\documentclass[11pt,a4paper]{article}

\usepackage{jcappub}
\usepackage{graphicx}
\usepackage{bm}
\usepackage{latexsym}
\usepackage{psfrag}
\usepackage{ulem}
\usepackage{textcomp}
\usepackage{color}
\usepackage{pstricks}

\makeatletter
\gdef\@fpheader{}
\makeatother

\def\benu{\begin{enumerate}}
\def\eenu{\end{enumerate}}
\def\nn{\nonumber} 

\def\f{\frac}
\def\l{\left}
\def\r{\right}
\def\d{{\rm d}}
\def\cR{{\mathcal R}}

\def\ei{\eta_{\rm i}}

\def\ee{\eta_{\rm e}}

\def\vx{{\bm x}}
\def\vk{{\bm k}}
\def\vka{{\bm k}_{1}}
\def\vkb{{\bm k}_{2}}
\def\vkc{{\bm k}_{3}}

\def\cB{{\cal B}}
\def\cG{{\cal G}}
\def\cI{{\cal I}}
\def\cJ{{\cal J}}

\def\fnl{f_{_{\rm NL}}}

\def\Mpl{M_{_{\rm Pl}}}
\def\Mp{M_{_{\rm Pl}}}

\newcommand{\bea}{\begin{eqnarray}}
\newcommand{\eea}{\end{eqnarray}}

\def\ps{{\mathcal P}_{_{\rm S}}}

\def\ns{n_{_{\rm S}}}

\def\cRB{{\cal R}^{_{^{^{\rm B}}}}}
\def\cRBe{{\cal R}^{_{{^{\rm B}}}}}

\begin{document}
\title{On the scalar consistency relation away from slow roll}
\author{V.~Sreenath$^\dag$,}
\affiliation{$^\dag$Department of Physics, Indian Institute of 
Technology Madras, Chennai~600036, India}
\emailAdd{sreenath@physics.iitm.ac.in}
\author{Dhiraj Kumar Hazra$^\ddag$}
\affiliation{$^\ddag$Asia Pacific Center for Theoretical Physics, Pohang, 
Gyeongbuk 790-784, Korea}
\emailAdd{dhiraj@apctp.org} 
\author{and L.~Sriramkumar$^\dag$}
\emailAdd{sriram@physics.iitm.ac.in}
\date{today} 
\abstract{
As is well known, the non-Gaussianity parameter $\fnl$, which is often used to
characterize the amplitude of the scalar bi-spectrum, can be expressed completely 
in terms of the scalar spectral index $n_{\rm s}$ in the squeezed limit, a 
relation that is referred to as the consistency condition.
This relation, while it is largely discussed in the context of slow roll 
inflation, is actually expected to hold in any single field model of 
inflation, irrespective of the dynamics of the underlying model, 
provided inflation occurs on the attractor at late times. 
In this work, we explicitly examine the validity of the consistency relation, 
analytically as well as numerically, {\it away from slow roll}.\/
Analytically, we first arrive at the relation in the simple case of power law 
inflation.
We also consider the non-trivial example of the Starobinsky model involving 
a linear potential with a sudden change in its slope (which leads to a brief
period of fast roll), and establish the condition completely analytically.
We then numerically examine the validity of the consistency relation in three 
inflationary models that lead to the following features in the scalar power 
spectrum due to departures from slow roll: (i)~a sharp cut off at large scales, 
(ii)~a burst of oscillations over an intermediate range of scales, and 
(iii)~small, but repeated, modulations extending over a wide range of scales. 
It is important to note that it is exactly such spectra that have been found 
to lead to an improved fit to the CMB data, when compared to the more standard 
power law primordial spectra, by the Planck team. 
We evaluate the scalar bi-spectrum for an arbitrary triangular configuration
of the wavenumbers in these inflationary models and explicitly illustrate that, 
in the squeezed limit, the consistency condition is indeed satisfied even in
situations consisting of strong deviations from slow roll. 
We conclude with a brief discussion of the results.}
\maketitle


\section{Introduction}\label{sec:introduction}

Over the past two decades, cosmologists have dedicated a considerable amount 
of attention to hunting down credible models of inflation. 
The inflationary scenario, which is often invoked to resolve certain puzzles 
(such as the horizon problem) that plague the hot big bang model, is well 
known to provide an attractive mechanism for the origin of perturbations 
in the early universe~\cite{oa,texts,reviews}. 
In the modern viewpoint, it is the primordial perturbations generated during
inflation that leave their signatures as anisotropies in the Cosmic Microwave 
Background (CMB) and later lead to the formation of the large scale structure. 
Ever since the discovery of the CMB anisotropies by COBE~\cite{cobe-1994}, there
has been a constant endeavor to utilize cosmological observations to arrive 
at stronger and stronger constraints on models of inflation.
While the CMB anisotropies have been measured with ever increasing precision 
by missions such as WMAP~\cite{wmap-2009,wmap-2011,wmap-2013},
Planck~\cite{planck-2013-cmbps,planck-2013-ccp,planck-2013-ci} and,
very recently, by BICEP2~\cite{bicep2-2014a,bicep2-2014b}, it would be fair 
to say that we still seem rather far from converging on a small class of well 
motivated and viable inflationary models (in this context, see 
Refs.~\cite{martin-2013,martin-2014a,martin-2014b}).

\par

The difficulty in arriving at a limited set of credible models of inflation 
seems to lie in the simplicity and efficiency of the inflationary scenario. 
Inflation can be easily achieved with the aid of one or more scalar fields 
that are slowly rolling down a relatively flat potential. 
Due to this reason, a plethora of models of inflation have been proposed,
which give rise to the required $60$ or so e-folds of accelerated expansion 
that is necessary to overcome the horizon problem. 
Moreover, there always seem to exist sufficient room to tweak the 
potential
parameters in such a way so as to result in a nearly scale invariant 
power spectrum of the scalar perturbations that lead to a good fit to 
the CMB data. 
In such a situation, non-Gaussianities in general and the scalar bi-spectrum 
in particular have been expected to lift the degeneracy prevailing amongst 
the various inflationary models.  
For convenience, the extent of non-Gaussianity associated with the scalar 
bi-spectrum is often expressed in terms of the parameter commonly referred 
to as $\fnl$~\cite{fnl-komatsu}, a quantity which is a dimensionless ratio 
of the scalar bi-spectrum to the power spectrum. 
The expectation regarding non-Gaussianities alluded to above has been largely 
corroborated by the strong limits that have been arrived at by the Planck 
mission on the value of the $\fnl$ parameter~\cite{planck-2013-cpng}. 
These bounds suggest that the observed perturbations are consistent with 
a Gaussian primordial distribution.  
Also, the strong constraints imply that exotic models which lead to large 
levels of non-Gaussianities are ruled out by the data.

\par 

Despite the strong bounds that have been arrived at on the amplitude of the 
scalar bi-spectrum, there exist many models of inflation that remain 
consistent with the cosmological data at hand.
The so-called scalar consistency relation is expected to play a powerful role 
in this regard, ruling out, for instance, many multi-field models of inflation, 
if it is confirmed observationally (for early discussion in this context, see, 
for instance, Refs.~\cite{maldacena-2003,creminelli-2004}; for recent discussions, 
see Refs.~\cite{cr-rd}; for similar results that involve the higher order 
correlation functions, see, for example, Refs.~\cite{npfs}).
According to the consistency condition, in the squeezed limit of the three-point 
functions wherein one of the wavenumbers associated with the perturbations is 
much smaller than the other two, the three-point functions can be completely 
expressed in terms of the two-point functions\footnote{It should be added
here that, in a fashion similar to that of the purely scalar case, one can also 
arrive at consistency conditions for the other three-point functions which 
involve tensors~\cite{tensor-bs,cc,cc2,sreenath-2013,kundu-2013,sreenath-2014}.}.
In the squeezed limit, for instance, the scalar non-Gaussianity parameter 
$\fnl$ can be expressed completely in terms of the scalar spectral index $\ns$ 
as $\fnl = 5\, (\ns -1)/12$~\cite{maldacena-2003,creminelli-2004}.
As we shall briefly outline later, the consistency conditions are expected 
to hold~\cite{sreenath-2014} whenever the amplitude of the perturbations 
freeze on super-Hubble scales, a behavior which is true in single field 
models where inflation occurs on the attractor at late times (see 
Refs.~\cite{oa}; in this context, also see Refs.~\cite{e-dfsr}).  
While the scalar consistency relation has been established in the slow roll 
scenario, we find that there has been only a limited effort in explicitly 
examining the relation in situations consisting of periods of fast 
roll~\cite{cr-d-d-sr-ar,cr-d-d-sr-nr}.
Moreover, it has been shown that there can be deviations from the consistency 
relation under certain conditions, particularly when the field is either 
evolving away from the attractor~\cite{dcr1} or when the perturbations are in 
an excited state above the Bunch-Davies vacuum~\cite{dcr2}.
In this work, our aim is to verify the validity of the scalar consistency 
relation in inflationary models which exhibit non-trivial dynamics. 
By considering a few examples, we shall explicitly show, analytically
and numerically, that the scalar consistency relation holds even in 
scenarios involving strong deviations from slow roll. 

\par

The remainder of this paper is organized as follows. 
In the next section, we shall quickly summarize a few essential points and
results concerning the scalar power spectrum and the bi-spectrum.
We shall also briefly revisit the proof of the scalar consistency relation 
in the squeezed limit.
In the succeeding section, we shall explicitly verify the validity of the
consistency condition analytically in the cases of power law inflation and 
the Starobinsky model which is described by a linear potential with a
sudden change in its slope.
We shall then evaluate the scalar bi-spectrum numerically for an arbitrary 
triangular configuration of the wavenumbers in three inflationary models 
that lead to features in the power spectrum, and examine the consistency
condition in the squeezed limit. 
We conclude with a brief discussion on the results we obtain.

\par

A few remarks on our conventions and notations seem essential at this stage
of our discussion.
We shall work with natural units wherein $\hbar=c=1$, and define the Planck 
mass to be $\Mp=(8\,\pi\, G)^{-1/2}$. 
We shall adopt the signature of the metric to be $(-,+,+,+)$. 
We shall assume the background to be the spatially flat, 
Friedmann-Lema\^itre-Robertson-Walker (FLRW) line element that is described by 
the scale factor $a$ and the Hubble parameter $H$. 
As is convenient, we shall switch between various parametrizations of time, 
viz. the cosmic time~$t$, the conformal time~$\eta$ or e-folds denoted by~$N$. 
An overdot and an overprime shall represent differentiation with respect to the
cosmic and the conformal time coordinates, respectively.
We shall restrict our attention in this work to inflationary models involving 
the canonical scalar field. 
Note that, in such a case, the first and second slow roll parameters are defined 
as $\epsilon_1 = -\dot H / H^2$ and $\epsilon_2 = \d \ln \,\epsilon_1/\d N$.  


\section{The scalar bi-spectrum in the squeezed limit}\label{sec:2}

In this section, we shall quickly summarize the essential definitions and
governing expressions concerning the scalar power spectrum, the bi-spectrum 
and the corresponding non-Gaussianity parameter~$\fnl$. 
We shall also sketch a simple proof of the consistency relation obeyed by 
the non-Gaussianity parameter $\fnl$ in the squeezed limit of the scalar 
bi-spectrum. 


\subsection{The scalar power spectrum and bi-spectrum}

Consider the following line-element which describes the spatially flat, 
FLRW spacetime, when the scalar perturbations, characterized by the 
curvature perturbation $\cR$, have been taken into account:
\begin{equation}\label{eq:metric}
\d s^2 = -\d t^2 
+ a^2(t)\; {\rm e}^{2\,{\cal R}(t,{\bm x})} \d {\bm x}^2.
\end{equation}
Let $f_k$ denote the Fourier modes associated with the curvature perturbation
at the linear order in the perturbations.
In the case of inflation driven by the canonical scalar field that is of our 
interest here, the modes $f_k$ satisfy the differential equation~\cite{texts,reviews}
\begin{eqnarray}\label{eq:ms}
f_k''+2\,\f{z'}{z}\, f_k'+k^2\,f_k = 0,\label{eq:de-fk}
\end{eqnarray}
where $z=\sqrt{2\, \epsilon_1}\, \Mp\, a$.
Upon quantization, the curvature perturbation can be decomposed in terms of 
the Fourier modes $f_k$ as 
\begin{eqnarray}
{\hat \cR}(\eta ,\vx)
=\int \f{{\rm d}^{3}{\vk}}{(2\,\pi)^{3/2}}\, 
{\hat {\cal R}}_{\vk}(\eta)\; {\rm e}^{i\, \vk \cdot \vx}
=\int \f{\d^{3}{\vk}}{(2\,\pi)^{3/2}}\, 
\l[{\hat a}_{\vk}\, f_{k}(\eta)\, {\rm e}^{i\,\vk\cdot \vx}
+{\hat a}_{\vk}^{\dagger}\, f_k^{\ast}(\eta)\, 
{\rm e}^{-i\,\vk \cdot \vx}\r],\label{eq:cR-d}
\end{eqnarray}
where ${\hat a}_{\vk}$ and ${\hat a}_{\vk}^{\dagger}$ are the usual creation 
and annihilation operators that obey the standard commutation relations. 

\par

The scalar power spectrum ${\cal P}_{_{\rm S}}(k)$ is defined in terms of the 
two-point correlation function of the curvature perturbation as follows:
\begin{equation}
\label{eq:ps}
\langle 0\vert {\hat \cR}_{\vk}(\eta)\, 
{\hat \cR}_{\vk'}(\eta)\vert 0\rangle 
=\f{(2\, \pi)^2}{2\, k^3}\; {\cal P}_{_{\rm S}}(k)\;
\delta^{(3)}\l(\vk+ \vk'\r),
\end{equation}
where $\vert 0\rangle$ denotes the Bunch-Davies vacuum annihilated by the 
operator ${\hat a}_{\vk}$~\cite{bunch-1978}.
In terms of the modes $f_k$, the scalar power spectrum 
${\cal P}_{_{\rm S}}(k)$ is given by
\begin{equation}
{\cal P}_{_{\mathrm{S}}}(k)
=\frac{k^{3}}{2\, \pi^{2}}\, \vert f_{k}\vert^{2}.
\label{eq:sps-d}
\end{equation}
The inflationary model governed by a given potential determines the behavior
of the quantity~$z$.
In order to arrive at the scalar power spectrum using the above expression, 
one first solves the differential equation~(\ref{eq:de-fk}) for the modes 
$f_k$ with the Bunch-Davies initial conditions~\cite{bunch-1978}, and then 
evaluates the amplitude of the modes at sufficiently late times when they 
are well outside the Hubble radius during inflation.
The scalar spectral index is defined as
\begin{equation}
\ns(k) = 1+\f{\d \ln \ps(k)}{\d \ln k}. \label{eq:ns}
\end{equation}
It should be stressed here that the scalar spectral index proves to be 
a constant only in simple situations such as power law and slow roll 
inflation.
In general, when the power spectrum contains features, the quantity 
$\ns$ depends on the wavenumber $k$. 

\par

The scalar bi-spectrum $\cB_{_{\rm S}}(\vka,\vkb,\vkc)$ evaluated, say,
in the vacuum state $\vert 0\rangle$, is defined in terms of the 
three-point function of the curvature perturbation as follows~\cite{wmap-2011}: 
\begin{equation}
\langle 0\vert {\hat \cR}_{\vka}\, {\hat \cR}_{\vkb}\, {\hat \cR}_{\vkc}
\vert 0 \rangle 
=\l(2\,\pi\r)^3\, \cB_{_{\rm S}}(\vka,\vkb,\vkc)\;
\delta^{(3)}\l(\vka+\vkb+\vkc\r).\label{eq:bi-s}
\end{equation}
Note that the delta function on the right hand side imposes the triangularity 
condition, viz. that the three wavevectors $\vka$, $\vkb$ and $\vkc$ have to 
form the edges of a triangle.   
For the sake of convenience, we shall set 
\begin{equation}
\cB_{_{\mathrm{S}}}(\vka,\vkb,\vkc)=\l(2\,\pi\r)^{-9/2}\; G(\vka,\vkb,\vkc).
\end{equation}
The non-Gaussianity parameter~$\fnl$ that is often used to characterize the
extent of non-Gaussianity indicated by the bi-spectrum is introduced through 
the relation~\cite{fnl-komatsu}
\begin{equation}
\cR(\eta, {\bm x})=\cR_{_{\mathrm{G}}}(\eta, {\bm x})
-\frac{3\,\fnl}{5}\, 
\l[\cR_{_{\mathrm{G}}}^2(\eta,{\bm x})
-\l\langle\cR_{_{\mathrm{G}}}^2(\eta, {\bm x})\r\rangle\r],
\label{eq:fnl-i}
\end{equation}
where $\cR_{_{\mathrm{G}}}$ denotes the Gaussian part of the curvature 
perturbation. 
Using the above relation and Wick's theorem (which applies to Gaussian 
perturbations), one can arrive at the following expression for the
dimensionless non-Gaussianity parameter $\fnl$ in terms of the bi-spectrum 
$G(\vka,\vkb,\vkc)$ and the scalar power spectrum ${\cal P}_{_{\mathrm{S}}}(k)$:
\begin{eqnarray}
\fnl(\vka,\vkb,\vkc)
&=&-\frac{10}{3}\; \f{1}{(2\,\pi)^{4}}\;
\l(k_{1}\, k_{2}\,k_{3}\r)^3\;  G(\vka,\vkb,\vkc)\nn\\
& &\times\l[k_1^{3}\; {\cal P}_{_{\rm S}}(k_2)\; {\cal P}_{_{\rm S}}(k_3)
+{\rm two~permutations}\r]^{-1}.\label{eq:fnl-d}
\end{eqnarray}

\par

The scalar bi-spectrum generated during inflation can be evaluated using the
Maldacena formalism~\cite{maldacena-2003}.
The approach basically makes use of the third order action governing the 
curvature perturbation and the standard rules of perturbative quantum field 
theory to arrive at the scalar three-point 
function~\cite{maldacena-2003,ng-ncsf,ng-reviews}. 
It is found that, in the case of inflation driven by the canonical scalar field,
the third order action consists of six terms and the scalar bi-spectrum receives
a contribution from each of these `vertices'.     
In fact, there also occurs a seventh term which arises due to a field 
redefinition, a procedure which is necessary to reduce the action to 
a simpler form. 
One can show that the complete contribution to the scalar bi-spectrum in the
perturbative vacuum can be written as~\cite{martin-2012a,hazra-2012,hazra-2013}
\begin{eqnarray}
G(\vka,\vkb,\vkc)
&\equiv & \sum_{C=1}^{7}\; G_{_{C}}(\vka,\vkb,\vkc)\nn\\
&\equiv & \Mp^2\; \sum_{C=1}^{6}\; 
\Biggl\{\l[f_{k_1}(\ee)\, f_{k_2}(\ee)\,f_{k_3}(\ee)\r]\; 
\cG_{_{C}}(\vka,\vkb,\vkc)\nn\\ 
& &+\l[f_{k_1}^{\ast}(\ee)\, f_{k_2}^{\ast}(\ee)\,f_{k_3}^{\ast}(\ee)\r]\;
\cG_{_{C}}^{\ast}(\vka,\vkb,\vkc)\Biggr\}
\nonumber \\ & &
+\, G_{7}(\vka,\vkb,\vkc),\label{eq:G}
\end{eqnarray}
where $f_k$ are the Fourier modes in terms of which we had decomposed the
curvature perturbation at the linear order in the perturbations
[cf.~Eq.~(\ref{eq:cR-d})].  
In the above expression, the  quantities $\cG_{_{C}}(\vka,\vkb,\vkc)$ with 
$C =(1,6)$ correspond to the six vertices in the interaction Hamiltonian
(obtained from the third order action), and are described by the integrals
\begin{subequations}
\label{eq:cG}
\begin{eqnarray}
\cG_{1}(\vka,\vkb,\vkc)
&=&2\,i\,\int_{\ei}^{\ee} \d\eta\; a^2\, 
\epsilon_{1}^2\, \l(f_{k_1}^{\ast}\,f_{k_2}'^{\ast}\,
f_{k_3}'^{\ast}+{\rm two~permutations}\r),\label{eq:cG1}\\
\cG_{2}(\vka,\vkb,\vkc)
&=&-2\,i\;\l(\vka\cdot \vkb + {\rm two~permutations}\r)\;
\int_{\ei}^{\ee} \d\eta\; a^2\, 
\epsilon_{1}^2\, f_{k_1}^{\ast}\,f_{k_2}^{\ast}\,
f_{k_3}^{\ast},\label{eq:cG2}\\
\cG_{3}(\vka,\vkb,\vkc)
&=&-2\,i\,\int_{\ei}^{\ee} \d\eta\; a^2\,
\epsilon_{1}^2\; \Biggl[\l(\f{\vka\cdot\vkb}{k_2^2}\r)\,
f_{k_1}^{\ast}\,f_{k_2}'^{\ast}\, f_{k_3}'^{\ast} 
+ {\rm five~permutations}\Biggr],\label{eq:cG3}\quad\\
\cG_{4}(\vka,\vkb,\vkc)
&=&i\,\int_{\ei}^{\ee} \d\eta\; a^2\,\epsilon_{1}\,
\epsilon_{2}'\, \l(f_{k_1}^{\ast}\,f_{k_2}^{\ast}\,
f_{k_3}'^{\ast}+{\rm two~permutations}\r),\label{eq:cG4}\\
\cG_{5}(\vka,\vkb,\vkc)
&=&\frac{i}{2}\,\int_{\ei}^{\ee} \d\eta\; 
a^2\, \epsilon_{1}^{3}\; \Biggl[\l(\f{\vka\cdot\vkb}{k_2^2}\r)\,
f_{k_1}^{\ast}\,f_{k_2}'^{\ast}\, f_{k_3}'^{\ast}
+ {\rm five~permutations}\Biggr],\label{eq:cG5}\\
\cG_{6}(\vka,\vkb,\vkc) 
&=&\frac{i}{2}\,\int_{\ei}^{\ee} \d\eta\; a^2\, 
\epsilon_{1}^{3}\;
\Biggl\{\l[\f{k_1^2\,\l(\vkb\cdot\vkc\r)}{k_2^2\,k_3^2}\r]\, 
f_{k_1}^{\ast}\, f_{k_2}'^{\ast}\, f_{k_3}'^{\ast}
+ {\rm two~permutations}\Biggr\}.\label{eq:cG6}\qquad
\end{eqnarray}
\end{subequations}
These integrals are to be evaluated from a sufficiently early time, say, $\ei$, 
when the initial conditions are imposed on the modes until very late times, say, 
towards the end of inflation at $\ee$. 
The additional, seventh term $G_{7}(\vka,\vkb,\vkc)$ arises due to the field 
redefinition and its contribution to the bi-spectrum $G(\vka,\vkb,\vkc)$ is 
given by
\begin{equation}
G_{7}(\vka,\vkb,\vkc)
=\frac{\epsilon_{2}(\eta_{\rm e})}{2}\,
\l(\vert f_{k_1}(\eta_{\rm e})\vert^{2}\, 
\vert f_{k_2}(\eta_{\rm e})\vert^{2} 
+ {\rm two~permutations}\r).\label{eq:G7}
\end{equation} 


\subsection{The consistency relation}

The squeezed limit refers to the case wherein one of the wavenumbers of 
the triangular configuration vanishes, say, $k_{3} \to 0$, leading to 
$\vkb = -\vka$. 
Or, equivalently, one of the modes is assumed to possess a wavelength which 
is much larger than the other two. 
The long wavelength mode would be well outside the Hubble radius.
In models of inflation driven by a single scalar field, the amplitude of the 
curvature perturbation freezes on super-Hubble scales, provided 
the inflaton evolves on the attractor at late times~\cite{e-dfsr,dcr1}.
As a result, the long wavelength mode simply acts as a background as far as 
the other two modes are concerned. 
If $\cRB$ is the amplitude of the curvature perturbation associated with the 
long wavelength mode, then the unperturbed part of the original FLRW metric 
will be modified to 
\begin{equation}
\d s^2 = -\d t^2 + a^2(t)\, {\rm e}^{2\,\cRBe}\, \d {\bm x}^2.
\end{equation}
In other words, the effect of the long wavelength mode is to modify the scale 
factor locally, which is equivalent to a spatial transformation of the form 
${\bm x'} = \Lambda\, {\bm x}$, with the components of the matrix $\Lambda$ 
being given by $\Lambda_{ij}={\rm e}^{\cRBe}\, \delta_{ij}$.
Under such a transformation, the modes of the curvature perturbation transform 
as $\cR_\vk \to {\rm det}~(\Lambda^{-1})\, {\cal R}_{\Lambda^{-1} \vk} $.  
Further, we have $\vert\Lambda^{-1}\, \vk\vert=(1-\cRB)\; k$ and 
$\delta^{(3)}(\Lambda^{-1}\,{\bm k}_1+\Lambda^{-1}\,{\bm k}_2)
={\rm det}~(\Lambda)\; \delta^{(3)}({\bm k}_1+{\bm k}_2)$.
Utilizing these relations, the scalar two-point function can be written as 
\begin{eqnarray}
\langle \hat{\cR}_{\vka}\, \hat{\cR}_{\vkb} \rangle_{k} 
&=& \f{(2\,\pi)^2}{2\,k_1^3}\; \ps(k_1)\; \l[1 - (\ns-1)\, \cRB\r]\,
\delta^{(3)}(\vka + \vkb),
\end{eqnarray}
where the suffix $k$ on the two-point function indicates that the correlator 
has been evaluated in the presence of a long wavelength perturbation. 
Upon using the above expression for the scalar power spectrum, we can write 
the scalar bi-spectrum in the squeezed limit 
as~\cite{creminelli-2004,kundu-2013,sreenath-2014}
\begin{eqnarray}
\langle\, \hat{\cR}_{\vka}\, \hat{\cR}_{\vkb}\, 
\hat{\cR}_{\vkc}\, \rangle_{k_3} 
&\equiv&  \langle\, \langle\, \hat{\cal R}_{\vka}\, \hat{\cal R}_{\vkb}\, 
\rangle_{k_3}\; \hat{\cal R}_{\vkc}\, \rangle \nn \\
&=& -\,\f{(2\,\pi)^{5/2}}{4\, k_1^3\, k_3^3}\, \l(\ns - 1\r)\,
\ps(k_1)\, \ps(k_3)\; \delta^{3}(\vka + \vkb).
\end{eqnarray}
On making use of this expression for the scalar bi-spectrum in the squeezed
limit and the definition of the scalar power spectrum, one can immediately 
arrive at the consistency relation for $\fnl$, viz. that 
$\fnl = 5\,(\ns - 1)/12$~\cite{maldacena-2003,creminelli-2004,cr-rd}.


\section{Analytically examining the validity of the condition away 
from slow roll}\label{sec:3}

As was outlined in the previous section, the only requirement for the validity 
of the consistency relation is the existence of a unique clock during inflation. 
Hence, in principle, this relation should be valid for any single field model 
of inflation irrespective of the detailed dynamics, 
if the field is evolving on the attractor at late times. 
Therefore, it should be valid even away from slow roll. 
In this section, we shall analytically examine the validity of the consistency
condition in scenarios consisting of deviations from slow roll. 
After establishing the relation first in the simple case of power law inflation, 
we shall consider the Starobinsky model which involves a brief period of fast
roll.


\subsection{The simple example of power law inflation}\label{sec:3.1}

We shall first consider the case of power law inflation with no specific 
constraints on the power law index, so that the behavior of the scale 
factor can be far different from that of its behavior in slow roll inflation.
In power law inflation, the scale factor can be written as
\begin{equation}
a(\eta)=a_1\, \l(\f{\eta}{\eta_1}\r)^{\gamma + 1},\label{eq:a-p-law}
\end{equation}
where $a_1$ and $\eta_1$ are constants, and $\gamma <-2$. 
In such a background, the Fourier modes $f_k$ associated with the curvature 
perturbation that satisfy the Bunch-Davies initial conditions are found to 
be~\cite{hazra-2012,p-law}
\begin{equation} 
f_{k}(\eta) = \f{1}{\sqrt{2\,\epsilon_1}\,\Mp\, a(\eta)}\, 
\sqrt{\f{-\pi\, \eta}{4}}\; {\rm e}^{-i\,\pi\,\gamma/2}\;
H_{-(\gamma+1/2)}^{(1)}(-k\,\eta),\label{eq:fk-p-law}
\end{equation}
where the first slow roll parameter $\epsilon_1$ is a constant given by 
$\epsilon_1=(\gamma+2)/(\gamma+1)$.
Note that $H_\nu^{(1)}(x)$ denotes the Hankel function of the first 
kind~\cite{gradshteyn-2007}, while the scale factor $a(\eta)$ is 
given by Eq.~(\ref{eq:a-p-law}).
For real arguments, the Hankel functions of the first and the second kinds,
viz. $H_\nu^{(1)}(x)$ and $H_\nu^{(2)}(x)$, are complex conjugates of 
each other~\cite{gradshteyn-2007}.
Moreover, as $x\to 0$, the Hankel function has the following form
\begin{equation}
\lim_{x \rightarrow 0}\, 
H_{\nu}^{(1)}(x) 
= \f{i}{\pi\,\nu}\,
\l[\Gamma(1-\nu)\, {\rm e}^{-i\,\pi\,\nu}\,\l(\f{x}{2}\r)^{\nu}
- \Gamma(1+\nu)\, \l(\f{x}{2}\r)^{-\nu}\r].\label{eq:Hf-lt}
\end{equation}
Upon using this behavior, one can show that the corresponding scalar power 
spectrum, evaluated at late times, i.e. as $\eta\to -0$, is given by
\begin{equation}
\ps(k)= \f{1}{2\,\pi^3\, \Mpl^2\, \epsilon_1}\,
\l(\f{\vert\eta_1\vert^{\gamma+1}}{a_1}\r)^2\,
\biggl\vert \Gamma[-(\gamma+1/2)]\biggr\vert^2\; 
\l(\f{k}{2}\r)^{2\,(\gamma+2)}, \label{eq:ps-p-law}
\end{equation}
where $\Gamma(x)$ represents the Gamma function~\cite{gradshteyn-2007}.
The scalar spectral index corresponding to such a power spectrum is evidently 
a constant and can be easily determined to be $\ns = 2\,\gamma+5$. 
If the consistency condition is true, it would then imply that the scalar
non-Gaussianity parameter has the value $\fnl = 5\,(\gamma+2)/6$ in the 
squeezed limit. 

\par

Let us now evaluate the scalar bi-spectrum in the squeezed limit using the 
Maldacena formalism and illustrate that it indeed leads to the above 
consistency condition for $\fnl$. 
It should be clear that, in order to arrive at the complete scalar bi-spectrum, 
we first need to carry out the integrals~(\ref{eq:cG}) associated with the six 
vertices, calculate the corresponding contributions 
$G_{_{C}}(\vka,\vkb,\vkc)$ for $C=(1,6)$, and lastly add the 
contribution $G_7(\vka,\vkb,\vkc)$ [cf. Eq.~(\ref{eq:G7})] that arises due to 
the field redefinition.
However, since $\epsilon_1$ is a constant in power law inflation, the 
second slow roll parameter $\epsilon_2$ vanishes identically.
As a result, the contribution corresponding to the fourth term that 
is determined by the integral~(\ref{eq:cG4}) as well as the seventh 
term $G_7(\vka,\vkb,\vkc)$ prove to be zero.
Moreover, in the squeezed limit of our interest, i.e. as $k_3\to0$, the 
amplitude of the mode $f_{k_3}$ freezes and hence its time derivative
goes to zero.
Therefore, terms that are either multiplied by the wavenumber corresponding 
to the long wavelength mode or explicitly involve the time derivative of the 
long wavelength mode do not contribute, as both vanish in the squeezed limit. 
Due to these reasons, one finds that it is only the first and the second 
terms, determined by the integrals~(\ref{eq:cG1}) and (\ref{eq:cG2}), that 
contribute in power law inflation.
After an integration by parts, we find that, in the squeezed limit, these two 
integrals can be combined to be expressed as
\begin{equation}
\lim_{k_3\to 0}\, \l[\cG_{1}(\vk, -\vk , \vkc) 
+\cG_{2}(\vk, -\vk , \vkc)\r]
= \lim_{k_3\to 0}\, 2\, i \, \epsilon_1^2\,f_{k_3}^\ast\, 
\l[\l(a^2\, f_{k}'^\ast\, f_{k}^\ast \r)_{-\infty}^{0} 
+ 2 \, k^2 \, \int_{-\infty}^{0} \d \eta \, a^2 \, f_{k}^{\ast 2} \r], 
\label{eq:cG12}
\end{equation}
where we have set $\ei=-\infty$ and $\ee=0$.
One can show that the derivative $f_k'$ can be written as 
\begin{equation} 
f_{k}'(\eta) = \f{-k}{\sqrt{2\,\epsilon_1}\,\Mp\, a(\eta)}\,
\sqrt{\f{-\pi\,\eta}{4}}\; {\rm e}^{-i\,\pi\,\gamma/2}\, 
H_{-(\gamma+3/2)}^{(1)}(-k\,\eta).
\end{equation}
Therefore, upon using this expression for the derivative $f_k'$, the 
behavior~(\ref{eq:Hf-lt}), the following asymptotic form of the Hankel 
function 
\begin{equation}
\lim_{x \rightarrow \infty}\, H_{\nu}^{(1)}(x) 
= \sqrt{\f{2}{\pi\,x}}\, {\rm e}^{i\,(x - \pi\,\nu/2- \pi/4)},\\
\end{equation}
and the integral~\cite{gradshteyn-2007}
\begin{equation}
\int \d x\, x\, \l[ H_{\nu}^{(1)}(x) \r]^2 
= \f{x^2}{2}\, \l\{\l[H_{\nu}^{(1)}(x) \r]^2 
- H_{\nu-1}^{(1)}(x)\, H_{\nu+1}^{(1)}(x)\r\},
\end{equation}
we find that the bi-spectrum in the squeezed limit can be written as
\begin{eqnarray}
\lim_{k_3\to 0}\, k^3\, k_3^3\, G(\vk, -\vk , \vkc) 
=-8\, \pi^4\, (\gamma+2)\,  {\cal P}_{_{\rm S}}(k)\,{\cal P}_{_{\rm S}}(k_3).
\end{eqnarray}
This expression and the definition~(\ref{eq:fnl-d}) for the scalar non-Gaussianity 
parameter then leads to $\fnl = 5\,(\gamma+2)/6 $, which is the result suggested by 
the consistency relation. 
We should add here that such a result has been arrived at earlier using 
a slightly different approach (see the third reference in Refs.~\cite{cr-rd}).


\subsection{A non-trivial example involving the Starobinsky model}\label{sec:3.2}

The second example that we shall consider is the Starobinsky model. 
In the Starobinsky model, the inflaton rolls down a linear potential which 
changes its slope suddenly at a particular value of the scalar 
field~\cite{starobinsky-1992}. 
The governing potential is given by
\begin{equation} 
V(\phi)
=\left\{ \begin{array}{rcl}                  
V_0 + A_+\,(\phi -\phi_0) &{\rm for}& \phi > \phi_0,\\  
V_0 + A_-\,(\phi -\phi_0) &{\rm for}& \phi < \phi_0,
\end{array}\right.\label{eq:V-sm}
\end{equation}
where $V_0$, $A_+$, $A_-$ and $\phi_0$ are constants. 
An important aspect of the Starobinsky model is the assumption that it is
the constant~$V_0$ which dominates the value of the potential around~$\phi_0$. 
Due to this reason, the scale factor always remains rather close to that of 
de Sitter.
This in turn implies that the first slow roll parameter $\epsilon_1$ remains 
small throughout the domain of interest.
However, the discontinuity in the slope of the potential at $\phi_0$ causes
a transition to a brief period of fast roll before slow roll is restored at
late times.
One finds that the transition leads to large values for the second slow roll 
parameter $\epsilon_2$ and, importantly, the quantity $\dot{\epsilon}_2$ 
grows to be even larger, in fact, behaving as a Dirac delta function {\it at}\/ 
the transition.
As we shall discuss, it is this behavior that leads to the most 
important contribution to the scalar bi-spectrum in the 
model~\cite{martin-2012a,arroja-2011-2012,martin-2014}.

\par

Clearly, it would be convenient to divide the evolution of the background 
quantities and the perturbation variables into two phases, before and after 
the transition at $\phi_0$.
In what follows, we shall represent the various quantities corresponding to the 
epochs before and after the transition by a plus sign and a minus sign (in 
the super-script or sub-script, as is convenient), while the values of the 
quantities at the transition will be denoted by a zero.
Let us quickly list out the behavior of the different quantities which we shall
require to establish the consistency relation.

\par

The first slow roll parameter before and after the transition is found to 
be~\cite{starobinsky-1992,martin-2012a,arroja-2011-2012,martin-2014} 
\begin{subequations}
\begin{eqnarray}\label{eq:srp1}
\epsilon_{1+}(\eta) 
& \simeq & \f{A_+^2}{18\,\Mp^2\,H_0^4},\label{eq:e1p}\\ 
\epsilon_{1-}(\eta) 
& \simeq & \f{A_-^2}{18\,\Mp^2\,H_0^4}\,
\l[1-\f{\Delta A}{A_-}\,\l(\f{\eta}{\eta_0}\r)^3\r]^2,\label{eq:e1m}
\end{eqnarray}
\end{subequations}
where $\Delta A=A_--A_+$, $H_0$ is the Hubble parameter determined by the
relation $H_0^2\simeq V_0/(3\,\Mpl^2)$, and $\eta_0$ denotes the conformal 
time when the transition takes place. 
The second slow roll parameter is given by
\begin{subequations}
\begin{eqnarray}
\epsilon_{2+}(\eta) 
&=& 4\,\epsilon_{1+},\label{eq:e2p}\\
\epsilon_{2-}(\eta)  
&=& \f{6\,\Delta\,A}{A_-}\,\f{(\eta/\eta_0)^3}{1\, -\, 
(\Delta\,A/A_-)\, \l(\eta/\eta_0\r)^3} + 4\,\epsilon_{1-}.\label{eq:e2m}
\end{eqnarray}
\end{subequations}
In fact, to determine the modes associated with the scalar perturbations and 
to evaluate the dominant contribution to the scalar bi-spectrum, we shall also 
require the behavior of the quantity $\dot{\epsilon}_2$.
One can show that $\dot{\epsilon}_2$ can be expressed as
\begin{equation}
\dot{\epsilon}_2 
= -\f{2\, V_{\phi\,\phi}}{H} + 12\,H\,\epsilon_1 
-3\, H\, \epsilon_2 - 4\, H\, \epsilon_1^2 + 5\, H\, \epsilon_1
\epsilon_2- \f{H}{2}\, \epsilon_2^2,
\end{equation}
where $V_{\phi\phi}= \d^2V/\d\phi^2$, and it should be stressed
that this is an exact relation.
It should be clear that the first term in the above expression involving 
$V_{\phi\phi}$ will lead to a Dirac delta function due to the discontinuity
in the first derivative of the potential in the case of the Starobinsky model. 
Hence, the dominant contribution to $\dot{\epsilon}_2$ {\it at the transition}\/ 
can be written as~\cite{arroja-2011-2012,martin-2014}
\begin{equation}
\dot{\epsilon}_{2}^{\,0}
\simeq \f{2\, \Delta\, A}{H_0}\, \delta^{(1)}(\phi\, -\, \phi_0)
= \f{6\, \Delta\, A}{A_+\,a_0}\, \delta^{(1)}(\eta\, - \, \eta_0),
\label{eq:e2p0}
\end{equation}
where $a_0$ denotes the value of the scale factor when $\eta=\eta_0$ .
Post transition, the dominant contribution to $\dot{\epsilon}_2$ is 
found to be~\cite{martin-2012a}
\begin{equation}
\dot{\epsilon}_{2-}
\simeq -3\, H\, \epsilon_{2-} - \f{H}{2}\, \epsilon_{2-}^2 
\simeq -\f{18\,H_0\,\Delta\,A}{A_-}\, 
\f{\l(\eta/\eta_0\r)^3}{\l[1 - (\Delta\,A/A_-)\, \l(\eta/\eta_0\r)^3\r]^2}.
\label{eq:e2pm}
\end{equation}
 
\par

Due to the fact that the potential is linear and also since the first 
slow roll parameter remains small, the modes $f_k$ governing the curvature 
perturbation can be described by the conventional de Sitter modes to a 
good approximation before the transition.
For the same reasons, one finds that the scalar modes can be described 
by the de Sitter modes soon after the transition as well.
However, due to the transition, the modes after the transition are 
related by the Bogoliubov transformations to the modes before the
transition.
Therefore, the scalar mode and its time derivative before the transition 
can be written as~\cite{starobinsky-1992,martin-2012a,gsr,arroja-2011-2012,martin-2014}:
\begin{subequations}\label{eq:sm-bt}
\begin{eqnarray}
f_k^{+}(\eta)
&=&\frac{i\, H_0}{2\, \Mp\, \sqrt{{k^3}\,\epsilon_{1+}}}\,
\l(1+i\,k\,\eta\r)\,{\rm e}^{-i\,k\,\eta},\label{eq:fk-bt}\\
f_k^{+}{}'(\eta)
&=&\frac{i\, H_0}{2\, \Mp\, \sqrt{{k^3}\,\epsilon_{1+}}}\,
\l[\f{3\,\epsilon_{1+}}{\eta}\,
\l(1+i\,k\,\eta\r)
+k^2\,\eta\r] {\rm e}^{-i\,k\,\eta}.\label{eq:fkp-bt}
\end{eqnarray}
\end{subequations}
Whereas, the mode and its derivative after the transition can be expressed 
as follows:
\begin{subequations}\label{eq:sm-at}
\begin{eqnarray}
f_k^{-}(\eta)
&=&\frac{i\,H_0\,\alpha_k}{2\,\Mp\,\sqrt{{k^3}\,\epsilon_{1-}}}\,
\l(1+i\,k\,\eta\r)\, {\rm e}^{-i\,k\,\eta}
-\frac{i\,H_0\,\beta_k}{2\,\Mp\,\sqrt{{k^3}\,\epsilon_{1-}}}
\l(1-i\,k\,\eta\right)\, {\rm e}^{i\,k\,\eta},\label{eq:fk-at}\\
f_k^{-}{}'(\eta)
&=&\frac{i\,H_0\,\alpha_k}{2\,\Mp\,\sqrt{{k^3}\epsilon_{1-}}}
\l[\l(\epsilon_{1-}+\f{\epsilon_{2-}}{2}\r)\,\f{1}{\eta}\,\l(1+i\,k\,\eta\r)
+k^2\,\eta\r]\, {\rm e}^{-i\,k\,\eta}\nn\\
&-&\frac{i\,H_0\,\beta_k}{2\,\Mp\,\sqrt{{k^3}\epsilon_{1-}}}
\l[\l(\epsilon_{1-}+\f{\epsilon_{2-}}{2}\r)\,\f{1}{\eta}\,\l(1-i\,k\,\eta\right)
+k^2\,\eta\r]{\rm e}^{i\,k\,\eta},\label{eq:fkp-at}
\end{eqnarray}
\end{subequations}
with $\alpha_k$ and $\beta_k$ denoting the Bogoliubov coefficients.
Upon matching the above modes and their time derivatives at the transition, 
the Bogoliubov coefficients can be determined to be
\begin{subequations}
\begin{eqnarray}
\alpha_k 
&=& 1+\frac{3\,i\,\Delta A}{2\,A_{+}}\;\frac{k_0}{k}\,
\left(1+\frac{k_0^2}{k^2}\right),
\label{eq:alphak-sm}\\
\beta_k 
&=& -\frac{3\,i\,\Delta A}{2\,A_+}\;\f{k_0}{k}\,
\l(1+\frac{i\, k_0}{k}\r)^2\, {\rm e}^{2\,i\,k/k_{0}},
\label{eq:betak-sm}
\end{eqnarray}
\end{subequations}
where $k_0 = -1/\eta_0 = a_0\,H_0$ denotes the mode that leaves the Hubble
radius at the transition.
At late times, the scalar mode behaves as 
\begin{equation}
f_k^-(\eta_{\rm e})
=\frac{i\, H_0}{2\, \Mp\, \sqrt{{k^3}\,\epsilon_{1-}(\eta_{\rm e})}}\,
\l(\alpha_k-\beta_k\r),\label{eq:fk-lt}
\end{equation}
where $\epsilon_{1-}(\ee)=A_{-}^2/(18\, \Mpl^2\, H_0^4)$.
Therefore, the scalar power spectrum, evaluated as $\eta\to 0$, can be 
expressed as
\begin{eqnarray}
\ps(k) 
&=& \l(\f{H_0}{2\,\pi}\r)^2\,
\l(\f{3\,H_0^2}{A_-}\r)^2\, 
\vert \alpha_k - \beta_k \vert^2 \nn\\
&=& \l(\f{H_0}{2\,\pi}\r)^2\, 
\l(\f{3\,H_0^2}{A_-}\r)^2\, 
\l[\cI(k) + \cI_{\rm c}(k)\, \cos\l(\f{2\,k}{k_0}\r) 
+ \cI_{\rm s}(k)\, \sin\l(\f{2\,k}{k_0}\r)\r],\label{eq:sps-sm}
\end{eqnarray}
where the quantities $\cI(k)$, $\cI_{\rm c}(k)$ and $\cI_{\rm s}(k)$ are 
given by
\begin{subequations}
\begin{eqnarray}
\cI(k) 
&=& 1 + \f{9}{2}\, \l(\f{\Delta A}{A_+}\r)^2\, \l(\f{k_0}{k}\r)^2 
+ 9\, \l(\f{\Delta A}{A_+}\r)^2\, \l(\f{k_0}{k}\r)^4 
+ \f{9}{2}\, \l(\f{\Delta A}{A_+}\r)^2\, \l(\f{k_0}{k}\r)^6,\\
\cI_{\rm c}(k) 
&=& \f{3\,\Delta A}{2\,A_+}\, \l(\f{k_0}{k}\r)^2\, 
\l[\l(\f{3\,A_-}{A_+} - 7\r) 
- \f{3\,\Delta A}{A_+}\, \l(\f{k_0}{k}\r)^4\r],\\
\cI_{\rm s}(k) 
&=& -\f{3\, \Delta A}{A_+}\, \f{k_0}{k}\, 
\l[1 + \l(\f{3\,A_-}{A_+} - 4 \r)\, \l(\f{k_0}{k}\r)^2 
+ \f{3\,\Delta A}{A_+}\, \l(\f{k_0}{k}\r)^4 \right].
\end{eqnarray}
\end{subequations}
Note that, because of the features in the power spectrum, the corresponding
scalar spectral index $\ns$ depends on the wavenumber $k$, and is found to be
\begin{eqnarray}
\ns(k) &=&\f{1}{2}\;
\biggl[\cI(k) + \cI_{\rm c}(k)\, \cos\l(\f{2\,k}{k_0}\r) 
+ \cI_{\rm s}(k)\, \sin\l(\f{2\,k}{k_0}\r)\biggr]^{-1}\nn\\
& &\times\,\l[\cJ(k)
+ \cJ_{\rm c}(k)\, \cos \l(\f{2\,k}{k_0} \r) 
+ \, \cJ_{\rm s}(k) \, \sin \l(\f{2\,k}{k_0} \r)\r],
\end{eqnarray}
where $\cJ(k)$, $\cJ_{\rm c}(k)$ and $\cJ_{\rm s}(k)$ are given by
\begin{subequations}
\label{eqs:cJ}
\begin{eqnarray}
\cJ(k) 
&=& 2 - 9\,\l(\f{\Delta A}{A_+}\r)^2\, \l(\f{k_0}{k}\r)^2 
- 54\, \l(\f{\Delta A}{A_+}\r)^2\, \l(\f{k_0}{k}\r)^4 
- 45\, \l(\f{\Delta A}{A_+}\r)^2\, \l(\f{k_0}{k}\r)^6, \\
\cJ_{\rm c}(k) 
&=& -\f{3\,\Delta A}{A_+}\, 
\l[4 +\, \l(\f{15\,A_-}{A_+} - 23\r)\, \l(\f{k_0}{k} \r)^2 
+ \f{12\,\Delta A}{A_+}\, \l(\f{k_0}{k}\r)^4 
- \f{15\,\Delta A}{A_+}\, \l(\f{k_0}{k}\r)^6\r],\qquad\quad\;\\
\cJ_{\rm s}(k) 
&=& -\f{6\,\Delta A}{A_+}\, \f{k_0}{k}\, 
\l[\l(\f{3\,A_-}{A_+} -7\r) 
- 2\, \l(\f{3\,A_-}{A_+}-4\r)\, \l(\f{k_0}{k}\r)^2
- \f{15\,\Delta A}{A_+}\, \l(\f{k_0}{k}\r)^4 \right].
\end{eqnarray}
\end{subequations}
If the consistency condition is indeed satisfied, then the non-Gaussianity 
parameter, as predicted by the relation, would prove to be
\begin{eqnarray}
\fnl(k) 
&=& \f{5}{12}\,\l[\ns(k) - 1\r] \nn\\
&=& \f{5}{24}\,
\l[{\cI(k) + \cI_{\rm c}(k)\, \cos\l(\f{2\,k}{k_0}\r) 
+ \cI_{\rm s}(k)\, \sin\l(\f{2\,k}{k_0}\r)}\r]^{-1}\nn\\
& &\times\,\biggl\{\l[\cJ(k) - 2\, \cI(k)\r] \,
+ \l[\cJ_{\rm c}(k) - 2\, \cI_{\rm c}(k)\r]\, \cos \l(\f{2\,k}{k_0} \r) 
+ \l[\cJ_{\rm s}(k) - 2\, \cI_{\rm s}(k)\r]\, 
\sin \l(\f{2\,k}{k_0} \r)\biggr\}.\label{eq:cr-fnl-sm} \nn\\
\end{eqnarray}

\par

Let us now  examine whether we do arrive at the same result upon using the
Maldacena formalism to compute the scalar bi-spectrum. 
It is known that, when there exist deviations from slow roll, it is the
fourth vertex that leads to the most dominant contribution to the 
bi-spectrum.
In other words, we need to focus on the contribution~$G_4(\vka,\vkb,\vkc)$
that is governed by the integral~(\ref{eq:cG4}).
Notice that the integral involves the quantity $\epsilon_2'$.
In the Starobinsky model, at the level of approximation we are working in,
$\epsilon_{2+}=4\,\epsilon_{1+}$, with $\epsilon_1$ being a constant 
[cf.~Eqs.~(\ref{eq:e2p}) and~(\ref{eq:e1p})].
Hence, $\epsilon_2'$ as well as the integral $\cG_4(\vka,\vkb,\vkc)$ vanish 
during the initial slow roll phase, prior to the transition.
However, as we discussed above, due to the discontinuity at $\phi_0$, 
$\dot{\epsilon}_2$ is described by a delta function at the transition 
[cf.~Eq.~(\ref{eq:e2p0})], whereas, post transition, it is given by 
Eq.~(\ref{eq:e2pm}).
Since the mode $f_k$ and its derivative are continuous, the contribution 
due to the delta function at the transition can be easily evaluated using 
the modes $f_k^{+}$ and the corresponding derivative $f_k^{+}{}'$ 
[cf.~Eqs.~(\ref{eq:sm-bt})].
Since we are interested in the squeezed limit, the contribution at the 
transition can be written as 
\begin{equation}
\lim_{k_3\to 0} \cG_{4}^{0}(\vk,-\vk,\vkc)
= \lim_{k_3\to 0} \f{12\,i\, a_0^2\, \epsilon_{1+}\,\Delta\, A}{A_+}\,
\l[f_{k}^{+\ast}(\eta_0)\,f_{k}^{+\prime\ast}(\eta_0)\;
f_{k_3}^{\ast}(\eta_0)\r].\label{eq:cG4-at}
\end{equation}
The corresponding contribution to the bi-spectrum can be easily evaluated 
using the late time behavior~(\ref{eq:fk-lt}) of the mode $f_k$.
The contribution after the transition is governed by the integral
\begin{equation}
\lim_{k_3\to 0} \cG_{4}^{-}(\vk,-\vk,\vkc)
=\lim_{k_3\to 0} 2\,i\, 
\int_{\eta_{0}}^{\ee} \d\eta\; a^2\,\epsilon_{1-}\,
\epsilon_{2-}'\, f_{k}^{-\ast}\,f_{k}^{-\prime\ast}\,
f_{k_3}^{-\ast}.\label{eq:cG4-m}
\end{equation}
We find that the resulting integral, arrived at upon making use of the 
behavior~(\ref{eq:e1m}) and~(\ref{eq:e2pm}) of the slow roll parameters 
and the modes~(\ref{eq:sm-at}), can be easily evaluated. 
On adding the above two contributions at the transition and post-transition, 
one can show that the bi-spectrum in the squeezed limit can be written as
\begin{eqnarray} 
\lim_{k_3\to 0} G(\vk,-\vk,\vkc)
&=&-\f{81\, H_0^{12}}{8\, A_+^2\, A_-^2}\,
\biggl\{\l[\cJ(k) - 2\, \cI(k)\r]\,
+ \l[\cJ_{\rm c}(k) - 2\, \cI_{\rm c}(k)\r]\, \cos \l(\f{2\,k}{k_0} \r)\nn\\ 
& &+\, \l[\cJ_{\rm s}(k) - 2\, \cI_{\rm s}(k)\r]\, 
\sin \l(\f{2\,k}{k_0} \r)\biggr\}.\label{eq:bs-sl-sm}
\end{eqnarray}

\par

There are a few points concerning this result that require emphasis.
The above bi-spectrum goes to a constant value at large scales, while it 
is found to oscillate with a constant amplitude in the small scale limit.
In the equilateral limit, the contribution at the transition is 
known to lead to a term that grows linearly with $k$ at large 
wavenumbers~\cite{arroja-2011-2012,martin-2014}. 
This essentially arises due to the infinitely sharp transition in the 
Starobinsky model. 
In the squeezed limit, one does not encounter such a growing term, but the 
sharpness of the transition is reflected in the oscillations of a fixed 
amplitude that persist indefinitely at small scales.
Clearly, one can expect these oscillations to die down at suitably large
wavenumbers if one smoothens the transition~\cite{martin-2014}.
As far as our primary concern here, viz. the validity of the consistency
condition, we find that, upon making use of the expression~(\ref{eq:bs-sl-sm}) 
for the bi-spectrum and the power spectrum~(\ref{eq:sps-sm}), we indeed 
recover the $\fnl$ as given by Eq.~(\ref{eq:cr-fnl-sm}), implying that the
consistency relation does hold even in the case of the infinitely sharp 
Starobinsky model.
Moreover, it is important to appreciate the point that, while it is the
contribution at the transition that dominates the amplitude of the 
non-Gaussianity parameter at large wavenumbers, the contribution 
after the transition proves to be essential for establishing the 
consistency relation at small wavenumbers.
This suggests that the contributions after the transition are essential 
in order to arrive at the complete bi-spectrum in the Starobinsky 
model~\cite{martin-2012a,martin-2014}.


\section{Numerical verification of the relation during deviations 
from slow roll}\label{sec:3.3}

In this section, we shall numerically examine the validity of the consistency
relation in three models that lead to features in the scalar power spectrum 
due to deviations from slow roll. 
We shall consider models that result in features of the following types:
(i)~a sharp drop in power at large scales, roughly associated with the
Hubble scale today (see Refs.~\cite{pi}; for recent discussions, see
Refs.~\cite{lp-ls}), (ii)~a burst of oscillations around scales 
corresponding to the multipoles of 
$\ell\simeq 20$--$40$~\cite{l-22-40,ci,hazra-2010,benetti-2011-13}, and (iii)~small 
and repeated modulations extending over a wide range of 
scales~\cite{pso,pahud-2009,flauger-2010,aich-2013,meerburg-2013-14,easther-2013}.
Such features are known to result in a better fit to the cosmological data than
the more simple and conventional, nearly scale invariant, spectra.
It should be highlighted that it is essentially these three types of spectra 
that have been considered by the Planck team while examining the possibility
of features in the primordial spectrum~\cite{planck-2013-ci}.
We should also clarify that, though the fit to the data improves in the presence 
of features, the Bayesian evidence does not necessarily alter significantly, as 
the improvement in the fit is typically achieved at the cost of a few extra 
parameters~\cite{planck-2013-ci,martin-2013,martin-2014a,martin-2014,be-fim}. 
Nevertheless, we believe that the possibility of features require to be explored 
further since repeated exercises towards model independent reconstruction of the 
primordial power spectrum seem to point to their presence~\cite{rc}.

\par

The different types of power spectra mentioned above can be generated by three 
inflationary models which we shall now briefly describe.
Power spectra with a sharp drop in power on large scales can be generated in 
scenarios dubbed punctuated inflation~\cite{pi}, which is a situation wherein 
a short period of departure from inflation is sandwiched between two epochs 
of slow roll inflation.
Such a punctuated inflationary scenario can be produced, for example, by the 
following potential which contains a point of inflection:
\begin{equation}
V(\phi) = \f{m^2}{2}\,\phi^2 
-\f{\sqrt{2\,\lambda\,(n-1)}\, m}{n}\, \phi^n
+\f{\lambda}{4}\,\phi^{2\,(n-1)},
\end{equation}
where $n\ge 3$.
The point of inflection proves to be crucial to recover the second stage of
slow roll inflation, after inflation has been interrupted briefly.
(We should clarify that we shall restrict ourselves to the $n=3$ case in this
work.)
The second class of features wherein there arises a burst of oscillations over
an intermediate range of scales can be generated by introducing a step in a
potential that otherwise leads to slow roll inflation.
The step results in a brief period of fast roll, which leads to the oscillations
in the scalar power spectrum.
For instance, if a step is introduced in the conventional quadratic potential,
the complete potential can be written as
\begin{equation}
V(\phi) = \f{m^2}{2}\,\phi^2\,
\l[1 + \alpha\,\tanh\,\l(\f{\phi -\phi_0}{\Delta\phi}\r)\r], 
\end{equation}
where, evidently, $\phi_0$, $\alpha$ and $\Delta\phi$ represent the location, 
the height and the width of the step, 
respectively~\cite{l-22-40,hazra-2010,benetti-2011-13}. 
Spectra with repeated modulations extending over a wide range of scales can be 
generated by potentials which contain oscillatory terms such as in the axion 
monodromy model~\cite{flauger-2010,aich-2013,easther-2013}.
The potential in such a case is given by
\begin{equation}
V(\phi) = \lambda\, 
\l[\phi + \alpha\, \cos\,\l(\f{\phi}{\beta} +\delta\r)\r].
\end{equation}
where $1/\beta$ represents the frequency of oscillations in the potential,
while $\delta$ is a phase.
(For the best fit values of the potential parameters, arrived at upon comparison 
with the CMB data, as well as for an illustration of the scalar power spectra that 
arise in the above three models, we would refer the reader to
Ref.~\cite{hazra-2013}.)
Let us now turn to the numerical evaluation of the scalar bi-spectrum in 
these models and the verification of the consistency relation.


\subsection{$\fnl$ for an arbitrary triangular configuration of the wavenumbers}

We shall make use of the code BI-spectra and Non-Gaussianity Operator, 
or simply, BINGO, which we had developed earlier, to calculate the scalar
bi-spectrum~\cite{hazra-2012}. 
BINGO is a Fortran~90 code that evaluates the scalar bi-spectrum in single 
field inflationary models involving the canonical scalar field.
It is based on the Maldacena formalism, and it efficiently computes all 
the various contributions to the bi-spectrum.
It should be clear from the Maldacena formalism that, in order to arrive at the 
scalar bi-spectrum, one first requires the behavior of the background quantities 
(such as, say, the scale factor and the slow roll parameters) and the scalar 
modes $f_k$.
Then, it is a matter of computing the various integrals that govern the scalar
bi-spectrum.
The evolution of the background quantities is arrived at by solving the equation
describing the scalar field.
Once we have the solution to the background, the scalar modes are obtained
by solving the corresponding differential equation, viz. Eq.~(\ref{eq:de-fk}),
with the standard Bunch-Davies initial conditions.
With these at hand, the integrals involved [cf.~Eqs.~(\ref{eq:cG})] can be 
carried out from a sufficiently early time to a suitably late time.  
In the context of power spectrum, it is well known that it is sufficient to
evolve the modes from a time when they are sufficiently inside the Hubble 
radius, say, from $k/(a\,H)=10^2$, till they are well outside, say, when
$k/(a\,H) = 10^{-5}$~\cite{ne-ps}.
One finds that, in order to arrive at the bi-spectrum, it suffices 
to carry out the integrals involved over roughly the same domain in 
time~\cite{hazra-2013,sreenath-2013,ng-ne,ng-f}.
However, two points need to be emphasized in this regard.
Firstly, in the case of the bi-spectrum, while evaluating for an arbitrary
triangular configuration, one needs to make sure that the integrals are 
carried out from a time when the largest of the three modes (in terms of
wavelength) is well inside the Hubble radius to a time when the smallest of 
the three is sufficiently outside.
To achieve the accuracy we desire (say, of the order of $2$--$3\%$ or better),
we perform the integrals from the time when the largest mode satisfies the
condition $k/(a\,H)=10^2$ until a time when the smallest mode satisfies the
condition $k/(a\,H) = 10^{-5}$.
(This is so barring the case of the axion monodromy model wherein we have to
integrate from deeper inside the Hubble radius---actually, from $k/(a\,H)=250$ 
for the values of the parameters that we work with---to take into account the 
resonances that occur in the model~\cite{flauger-2010}.)
Secondly, due to continued oscillations in the sub-Hubble domain, it is well 
known that the integrals require a cut-off in order for them to converge.
We have introduced a cut-off of the form $\exp\,[-\kappa\,k/(a\,H)]$ and have 
worked with $\kappa = 0.1$, which is known to lead to consistent 
results~\cite{hazra-2013,sreenath-2013}.
We should mention here that we have made the latest version of BINGO publicly 
available at the URL: {\tt http://www.physics.iitm.ac.in/\~{}sriram/bingo/bingo.html}.
The earlier public version of the code was limited to the evaluation of the 
bi-spectrum in the equilateral limit. 
The current version can compute the bi-spectrum for an arbitrary triangular
configuration of the wavenumbers, including the squeezed limit of our interest
here\footnote{We should add that we have independently reproduced the results 
being presented here using a different code as well.
The latter code was originally used to calculate scalar-tensor three-point
functions and the tensor bi-spectrum~\cite{sreenath-2013}, and it has been 
modified suitably to calculate the scalar bi-spectrum and the corresponding 
non-Gaussianity parameter $\fnl$.}.
 
\par

Before we go on to consider the consistency relation in the squeezed limit, 
let us make use of BINGO to understand the shape and structure of the 
bi-spectrum or, equivalently, the 
\begin{figure}[!ht]
\begin{center}
\begin{tabular}{cc}
\hskip -5pt
\includegraphics[width=10.00cm]{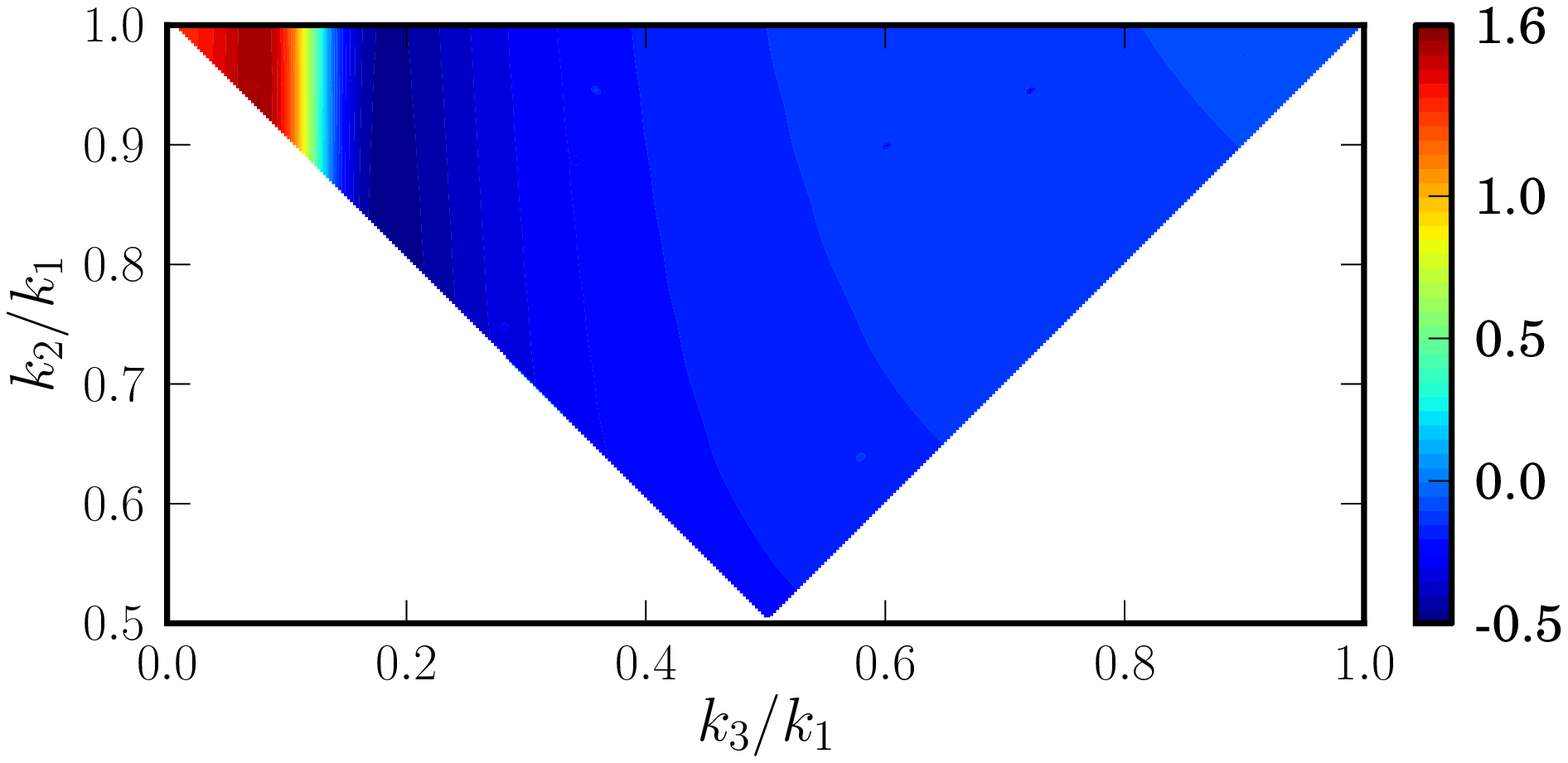}\\
\includegraphics[width=10.00cm]{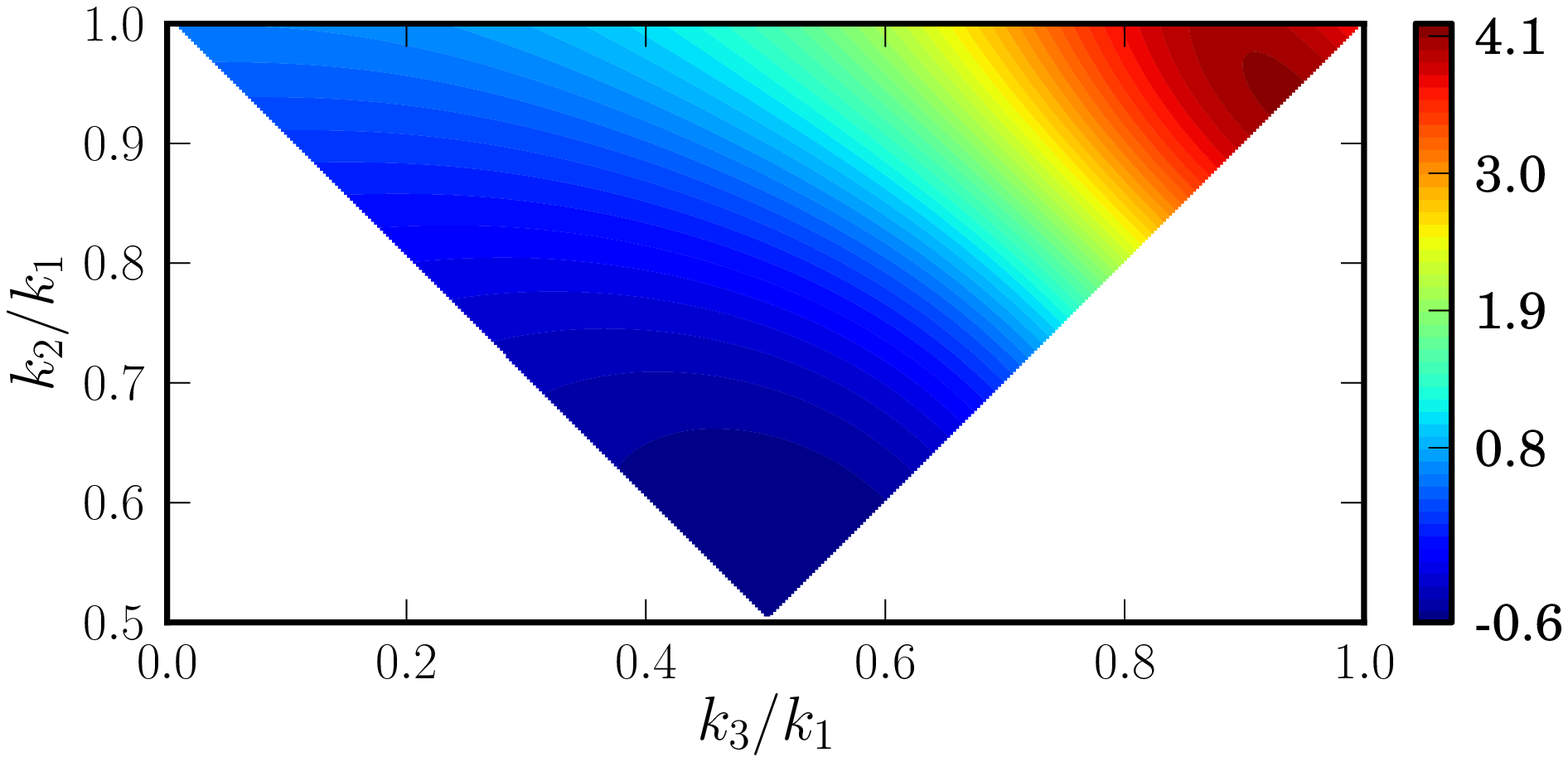}\\
\includegraphics[width=10.00cm]{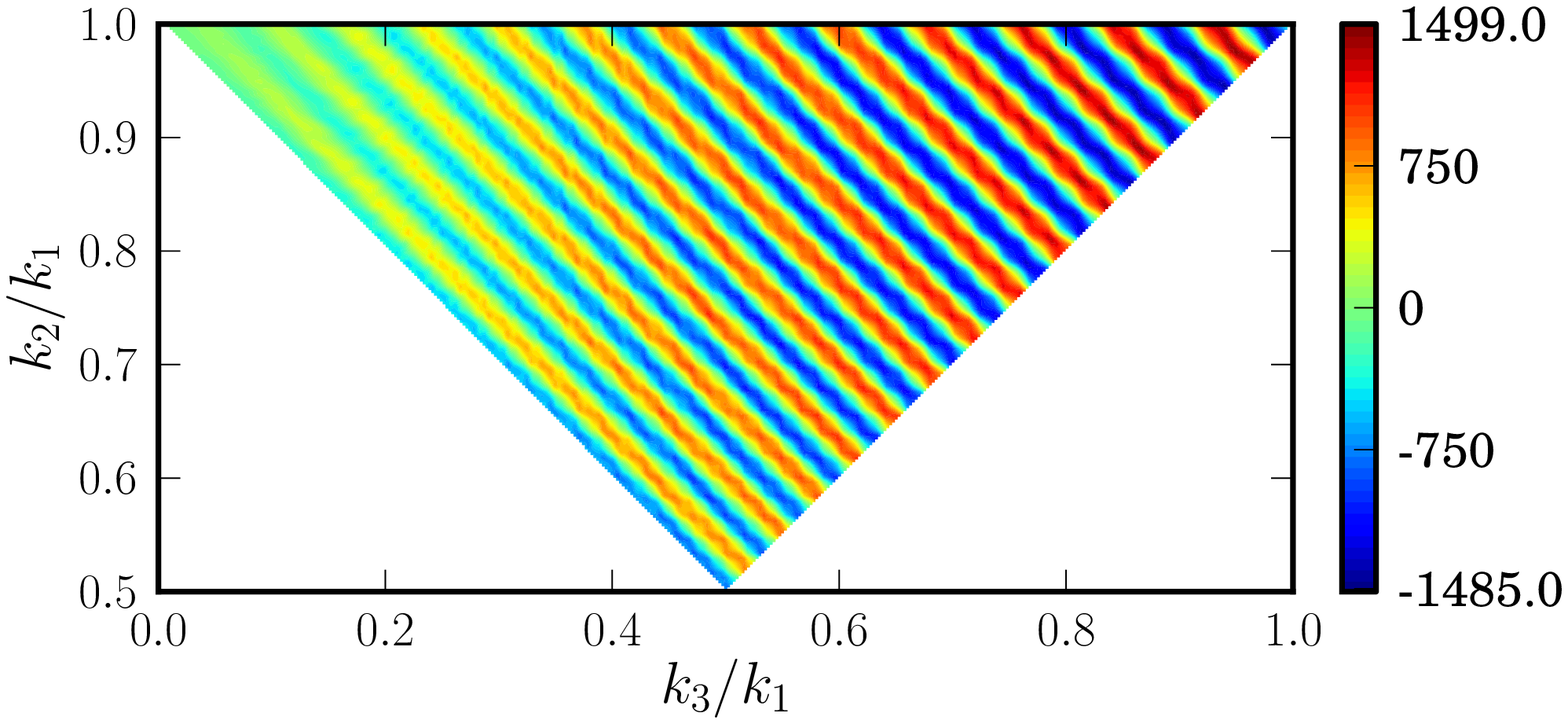}\\
\end{tabular}
\caption{\label{fig:bs-tri} Density plots of the scalar non-Gaussianity 
parameter $\fnl$, plotted as function of $k_3/k_1$ and $k_2/k_1$, with 
fixed values of $k_1$, in the three models of our interest, viz. 
punctuated inflation (on top), quadratic potential with a step (in the 
middle) and axion monodromy model (at the bottom). 
We have chosen the wavenumber $k_1$ to be $10^{-3}\; {\rm  Mpc}^{-1}$ in 
the case of punctuated inflation, while we have set it to be $2\times 10^{-3}\; 
{\rm Mpc}^{-1}$ for the other two models.
We find that, when the features are localized, as in the cases of punctuated 
inflation and the quadratic potential with a step, the structure of $\fnl$ 
varies considerably with the choice of $k_1$. 
However, in the case of the axion monodromy model, wherein there arises 
continued oscillations, the shape of $\fnl$ is more or less independent 
of the choice of $k_1$.}
\end{center}
\end{figure}
non-Gaussianity parameter $\fnl$, for an arbitrary triangular configuration of 
the wavenumbers\footnote{We should mention here that, apart from the scalar 
bi-spectrum, we shall also require the scalar power spectrum to arrive
at the non-Gaussianity parameter $\fnl$.
BINGO, as it computes the scalar modes, can easily be made use of to obtain
the power spectrum too.}.
Usually, the scalar bi-spectrum and the parameter $\fnl$ are illustrated as 
density plots, plotted as a function of the ratios $k_3/k_1$ and $k_2/k_1$, 
for a fixed value of $k_1$ (in this context, see, for instance, 
Ref.~\cite{hazra-2013}).
While the actual value of $k_1$ will not play a significant role in simple 
slow roll scenarios, the structure of the bi-spectrum revealed in such
density plots will depend on choice of $k_1$ in models which lead to 
features. 
In Fig.~\ref{fig:bs-tri}, we have plotted the scalar non-Gaussianity parameter
arising in the three inflationary models of our interest, for suitable values 
of the quantity $k_1$. 
We find that, in the cases of punctuated inflation and the quadratic potential 
with a step, since the features are localized over a small range of scales, the 
structure of the plot changes to a certain extent with the choice of $k_1$.
However, in the case of axion monodromy model, because of the reason that the 
oscillations extend over a wide range of scales, the choice of $k_1$ does not 
alter the structure of the plots significantly.

\par

\begin{figure}[!ht]
\begin{center}
\vskip -30pt
\hskip -15pt
\includegraphics[trim=0.0cm 6.0cm 0.0cm 6.0cm,clip=true,width=8.00cm]{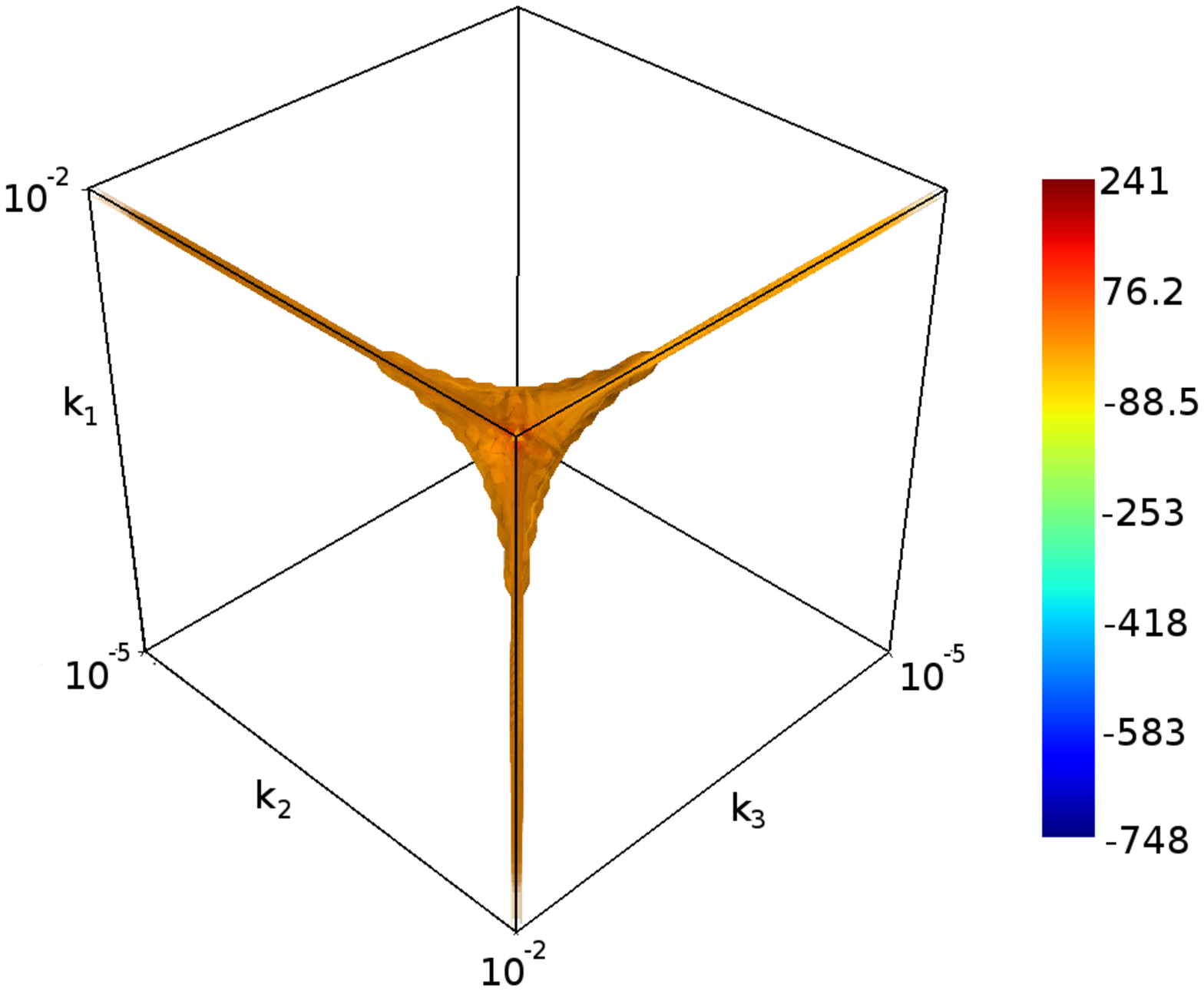} 
\hskip -10pt
\includegraphics[trim=0.0cm 6.0cm 0.0cm 6.0cm,clip=true,width=8.00cm]{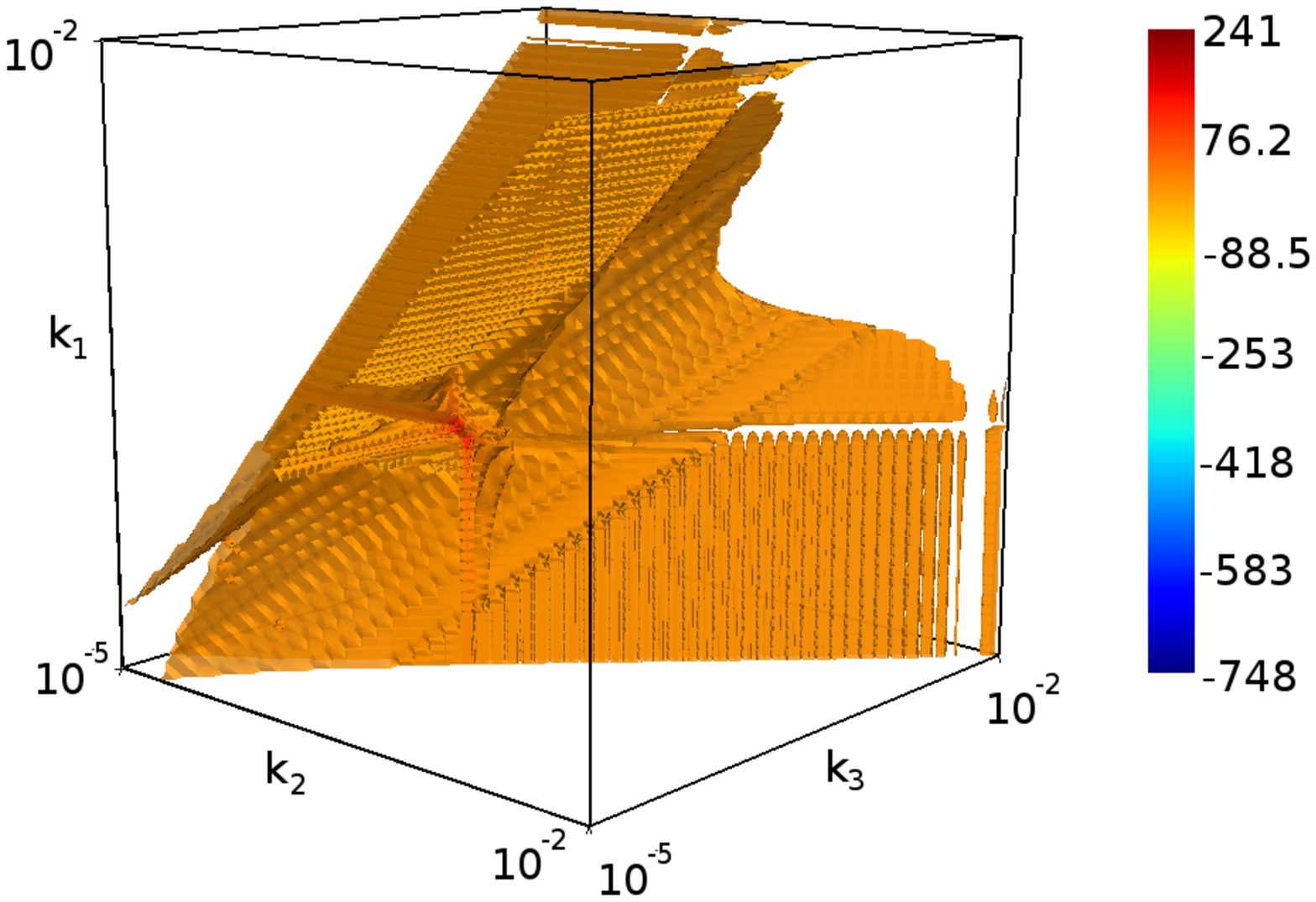}
\vskip -20pt
\hskip -15pt
\includegraphics[trim=0.0cm 6.0cm 0.0cm 6.0cm,clip=true,width=8.00cm]{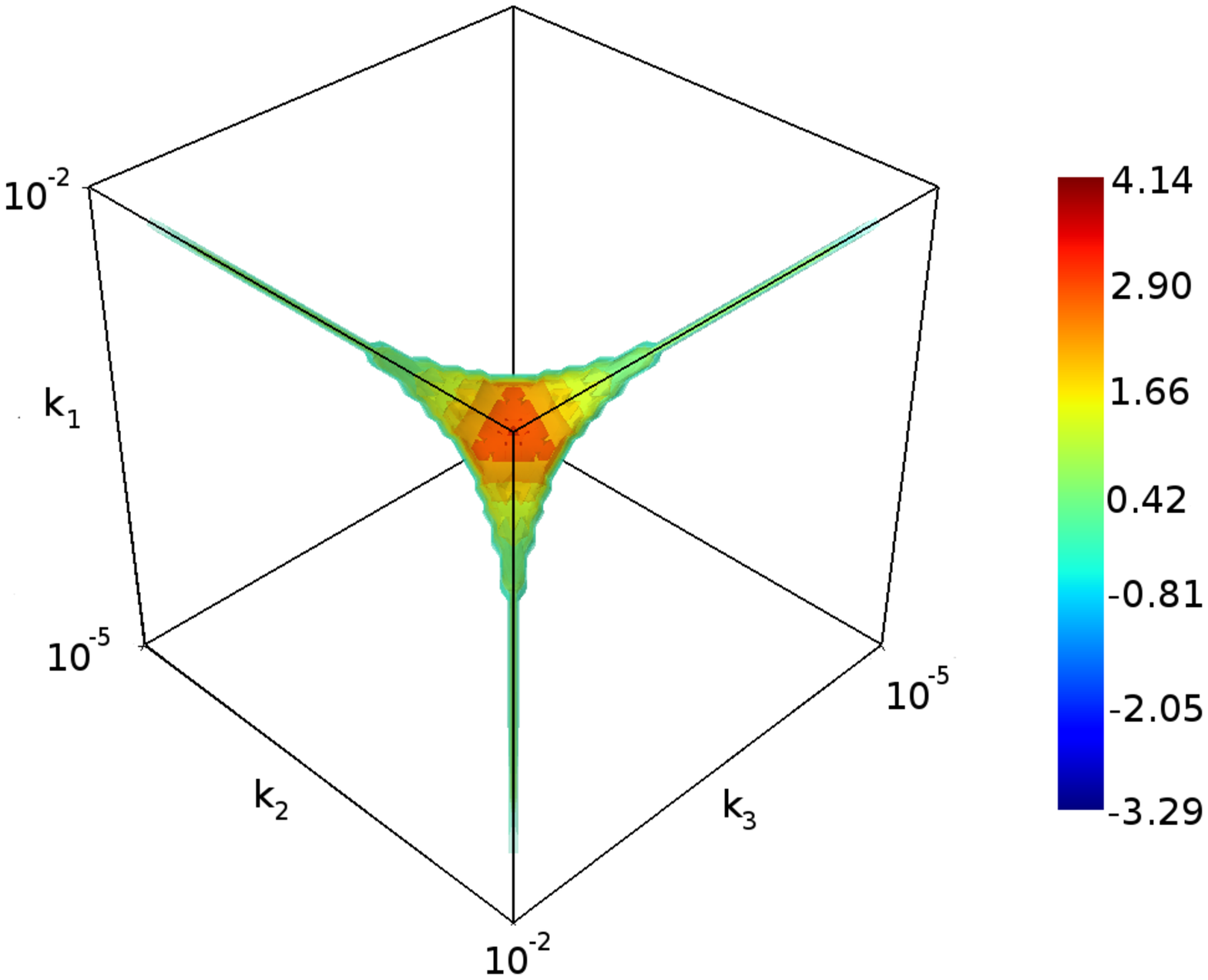} 
\hskip -10pt
\includegraphics[trim=0.0cm 6.0cm 0.0cm 6.0cm,clip=true,width=8.00cm]{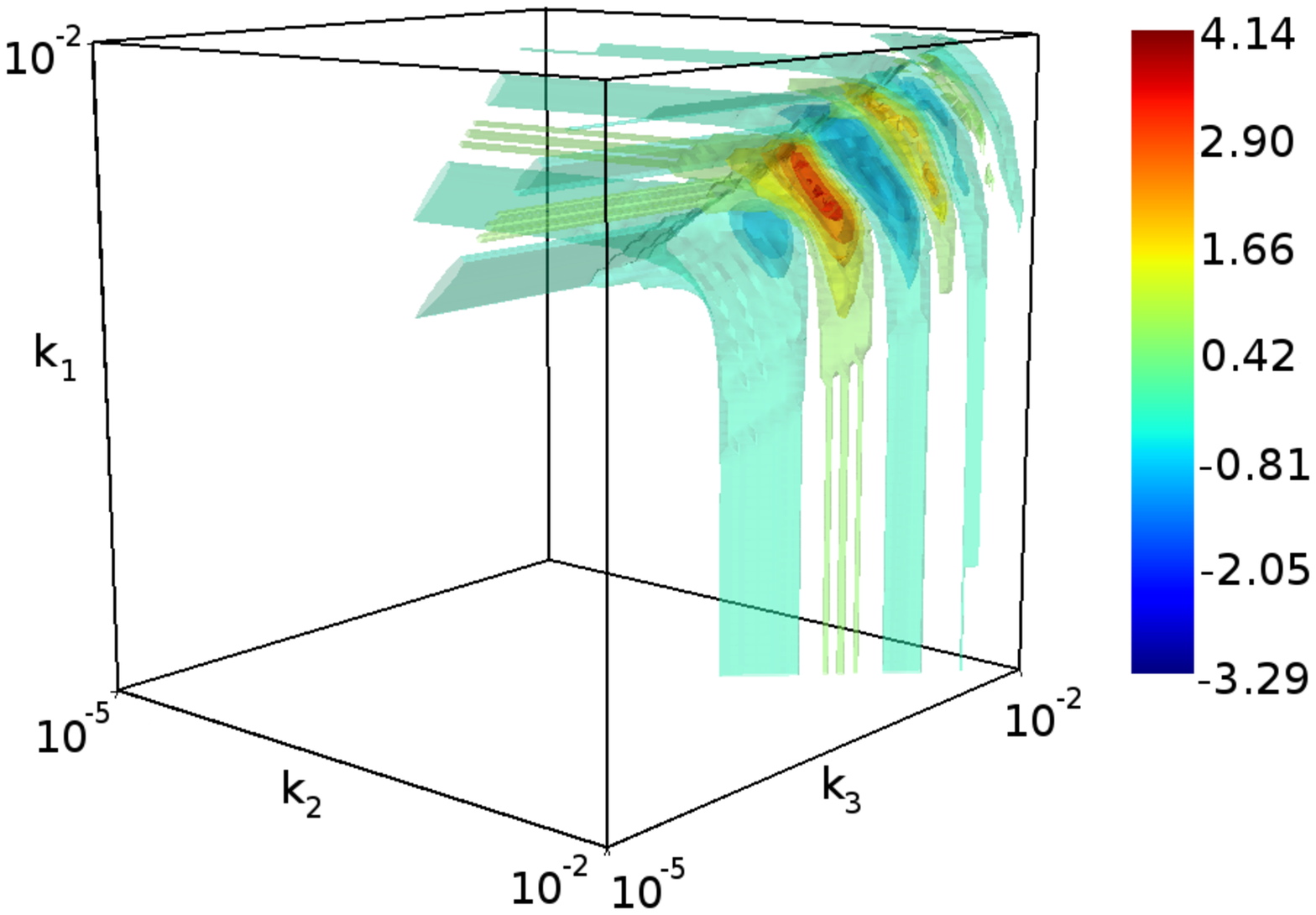}
\vskip -20pt
\hskip -15pt
\includegraphics[trim=0.0cm 6.0cm 0.0cm 6.0cm,clip=true,width=8.00cm]{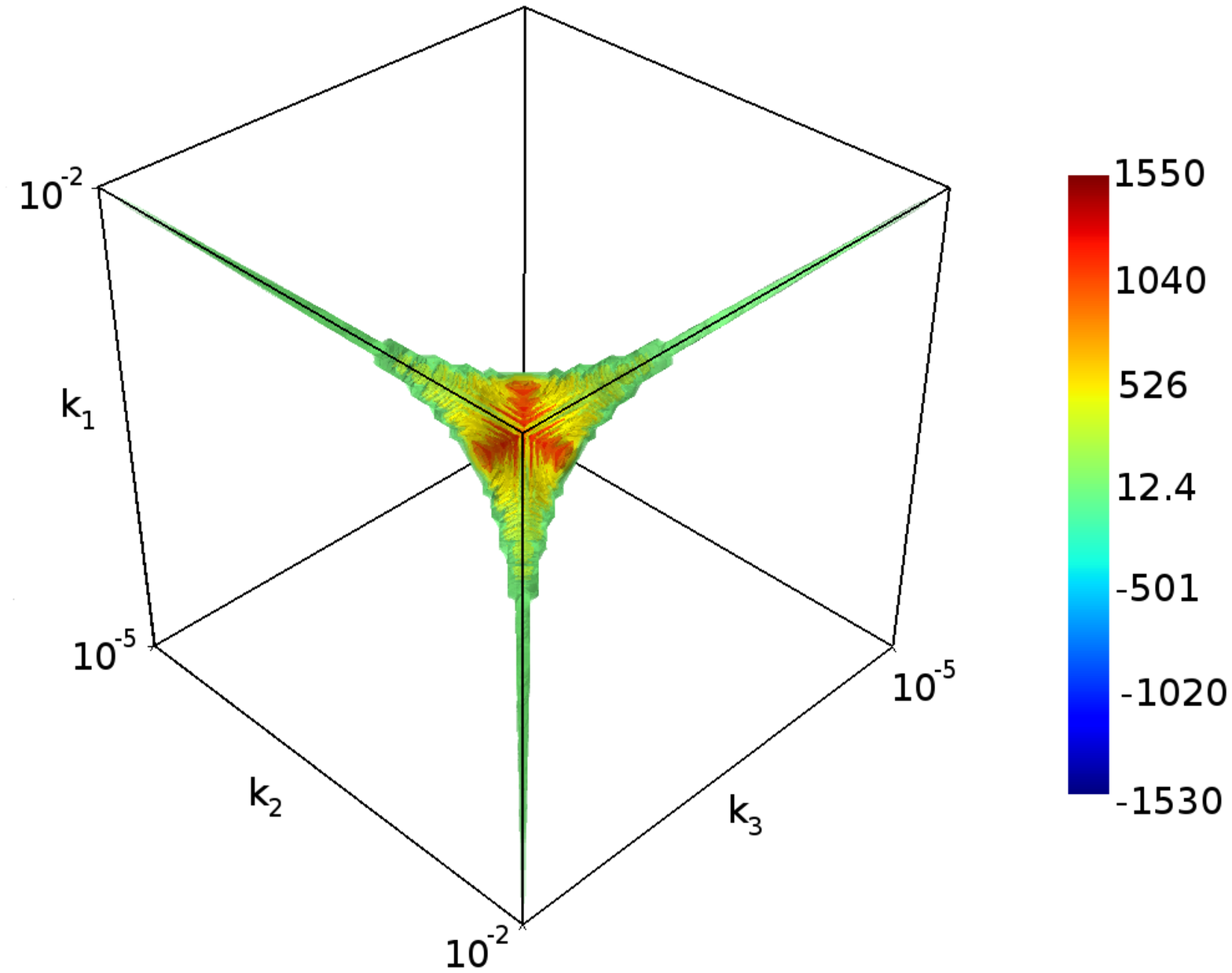} 
\hskip -10pt
\includegraphics[trim=0.0cm 6.0cm 0.0cm 6.0cm,clip=true,width=8.00cm]{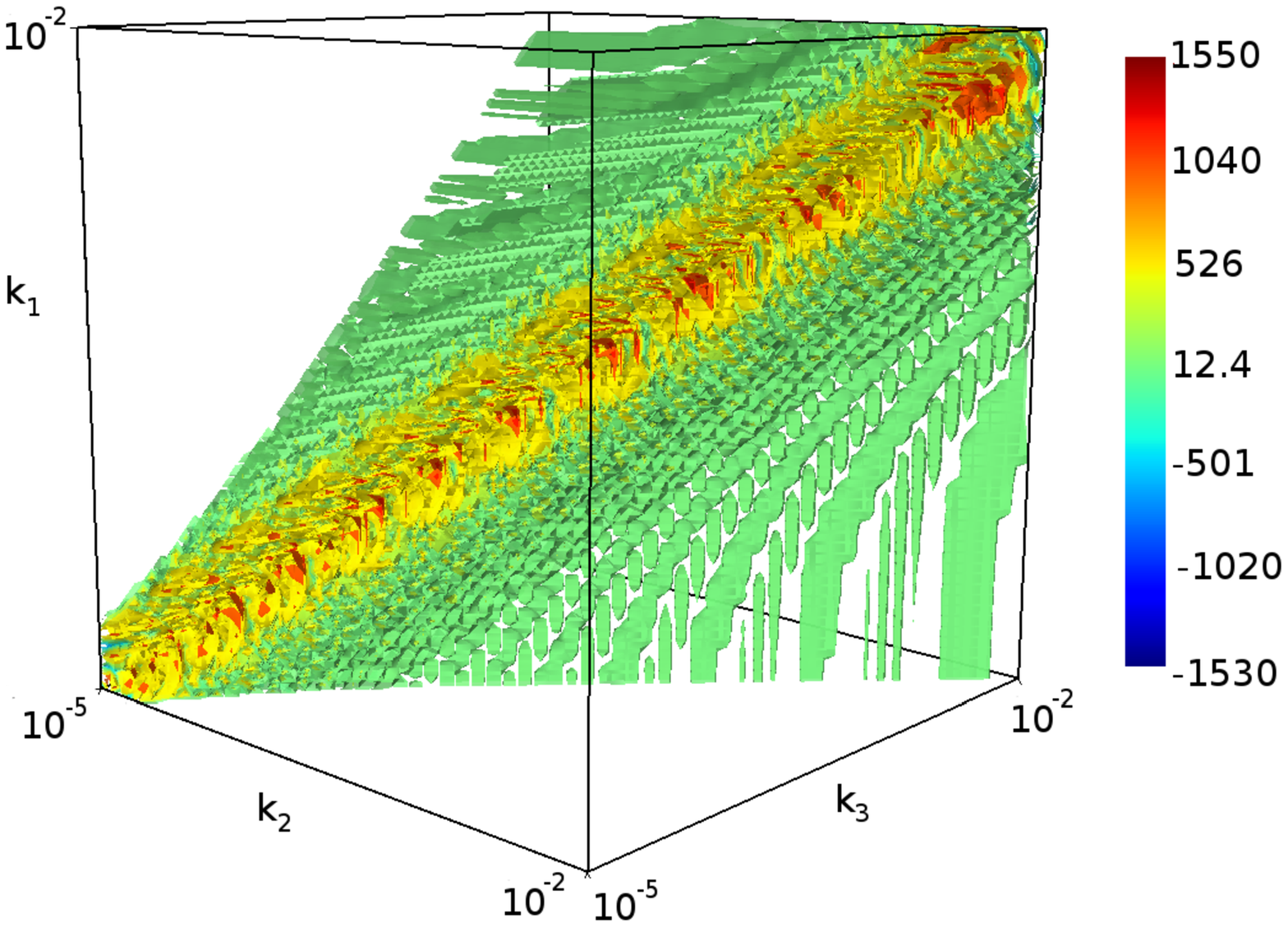}
\end{center}
\vskip -25pt
\caption{\label{fig:bs-t} Three-dimensional contour plots of the parameter 
$\fnl$, plotted against the three wavenumbers $k_1$, $k_2$ and $k_3$ in the 
three models of interest, organized in the same order as in the last
figure.
We have shown two different projections of the plots in the figure. 
The projections in the left column clearly indicate the symmetry along the 
three different axes, as is expected in an isotropic background.
The figures on the right column illustrate the fact that the allowed domain
of wavenumbers are confined to a `tetrapyd' and that the bi-spectrum peaks 
in the equilateral limit, i.e. along the line $k_1=k_2=k_3$.}
\end{figure}

In Fig.~\ref{fig:bs-t}, we have attempted to capture the complete structure 
and shape of the bi-spectrum using a three-dimensional contour plot.
We have made use of Mayavi and Python to create the three-dimensional 
plot~\cite{mayavi}.
We have plotted the parameter $\fnl$ for a wide range of the wavenumbers $k_1$, 
$k_2$ and $k_3$, over the allowed domain wherein the corresponding wavevectors 
satisfy the triangularity condition.
It is known that the triangularity condition restricts the wavenumbers to a
`tetrapyd', as is evident from the figure.  
In the figure, we have presented two projections of the three-dimensional plot. 
One of the views clearly shows the fact that the bi-spectrum is symmetric 
along the three axes, as is expected in an isotropic background.  
The second illustrates the fact the non-Gaussianity parameter peaks in the 
equilateral limit, i.e. when $k_1=k_2=k_3$.


\subsection{$\fnl$ in the squeezed limit}

\begin{figure}[!htbp]
\begin{center}
\begin{tabular}{c}
\hskip -10pt
\includegraphics[width=8.5cm]{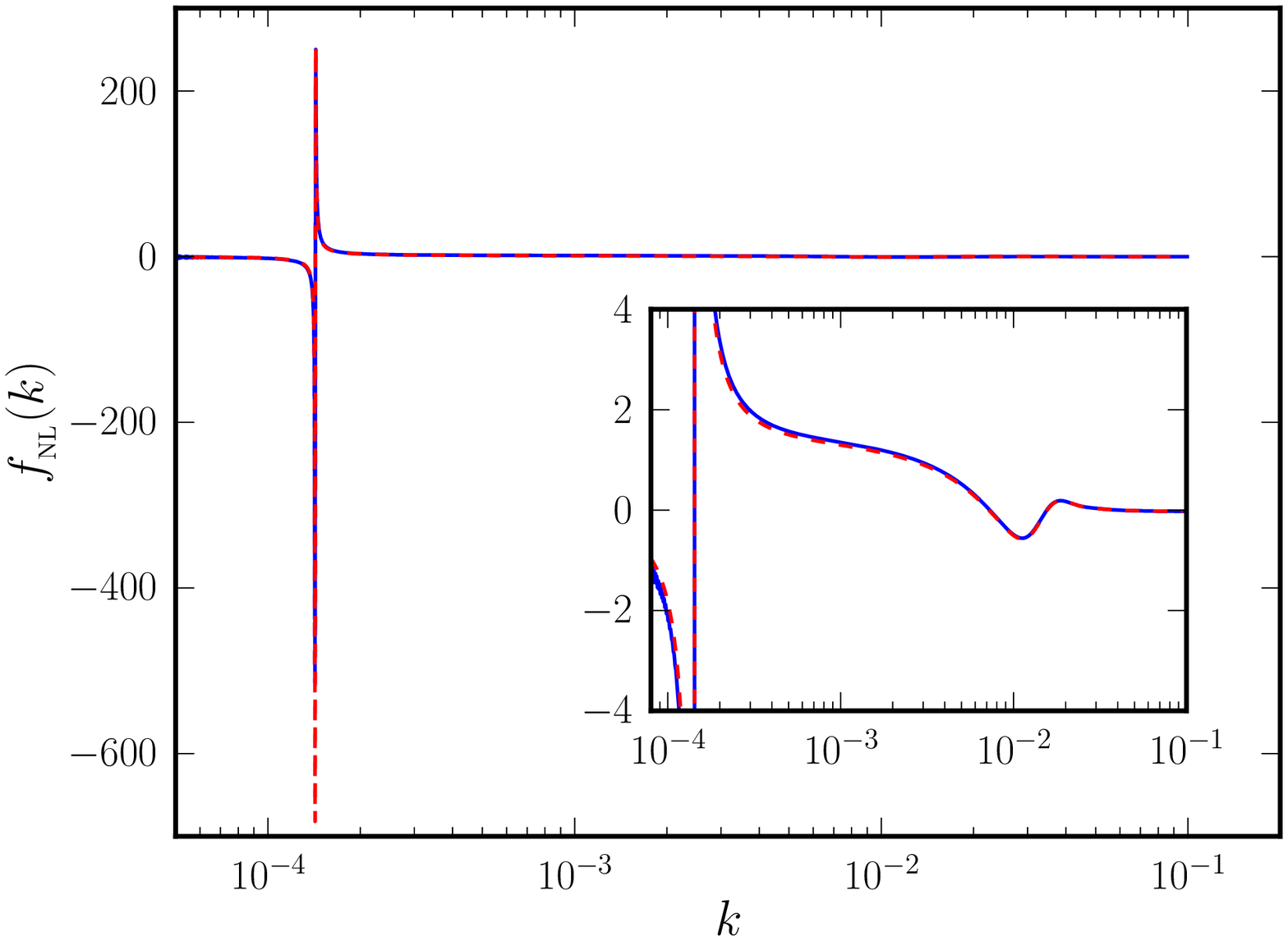} \\
\includegraphics[width=8.5cm]{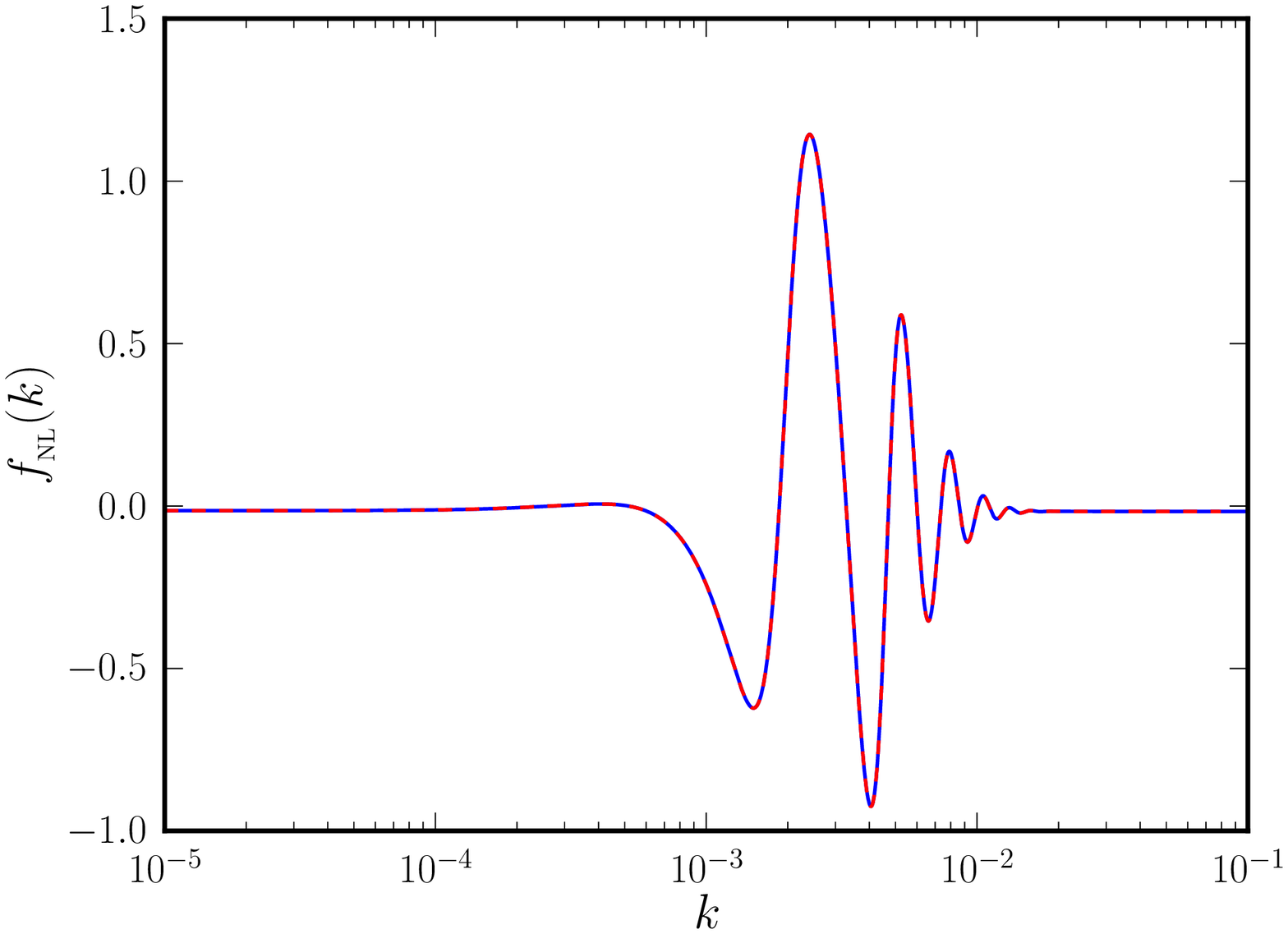} \\
\includegraphics[width=8.5cm]{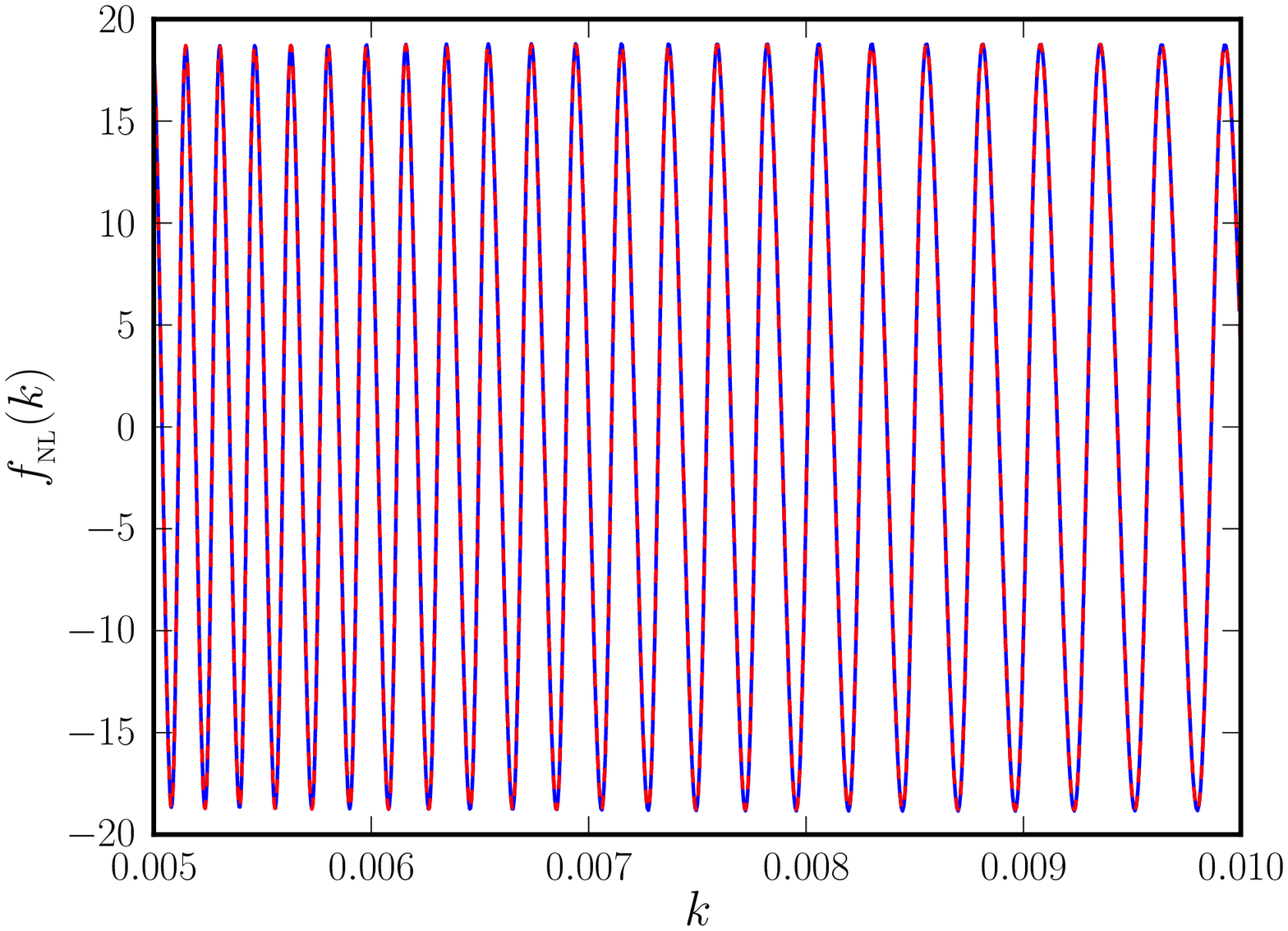} 
\end{tabular}
\caption{\label{fig:cr-pqa} The behavior of the non-Gaussianity parameter
$\fnl$ in the squeezed limit has been plotted as a function of $k$ for the
three models, organized as in the previous two figures. 
The blue curve represents the $\fnl$ calculated using Maldacena formalism, 
while the red dashed line corresponds to the same quantity arrived at from 
the consistency relation. 
The excellent match between the two curves indicate that the consistency 
relation is valid even in non-trivial scenarios involving brief departures 
from inflation.}
\end{center}
\end{figure}

Let us now turn to examine the consistency relation in the three models of 
our interest.
Towards this end, we have made use of BINGO to evaluate the non-Gaussianity
parameter $\fnl$ in the squeezed limit, using the Maldacena formalism.
As we had pointed out, BINGO can be made use of to evaluate the power spectrum
as well.
Using the expression~(\ref{eq:ns}) and the scalar power spectrum, we arrive at 
the scalar spectral index $\ns$, which we then utilize to verify the consistency
condition $\fnl (k) = 5\,\l[\ns(k)-1\r]/12$.
Before we go on to illustrate the results for the three models that we are
focusing on, a couple of points concerning the squeezed limit needs to 
be made.
We should stress that we take the wavenumber of the squeezed mode to be smallest 
wavenumber that is numerically tenable in the sense that the mode is sufficiently 
inside the Hubble radius at a time close to when the integration of the background 
begins.
Moreover, it should be noted that, since the squeezed mode has a finite and
non-zero wavenumber, in the squeezed limit of our interest, the numerically 
evaluated bi-spectrum is expected to be more accurate at larger wavenumbers
than the smaller ones.   
In Fig.~\ref{fig:cr-pqa}, we have plotted the quantity $\fnl$ obtained from the
Maldacena formalism as well as the quantity arrived at from the consistency 
relation.
It is clear that the two quantities match very well (they match at the level 
of a few percent) thus confirming the validity of consistency relation even 
in scenarios displaying highly non-trivial dynamics.

\par


\section{Discussion}\label{sec:d}

At the level of the three-point function, the consistency condition relates 
the scalar bi-spectrum to the power spectrum in the squeezed limit wherein
the wavelength of one of the three modes is much longer than the other two.
As we had discussed, the consistency condition applies to any situation
wherein the amplitude of the long wavelength mode freezes.
Since the amplitude of the curvature perturbation settles down to a constant 
value on super-Hubble scales in most single field models of inflation, the 
consistency relation is expected to be valid in such models.
It is easy to analytically establish the consistency relation in slow roll
scenarios.
In contrast, one needs to often resort to numerical methods to analyze 
situations involving departures from slow roll. 
In this work, we had examined the validity of the consistency condition, 
analytically as well as numerically, in a class of models permitting
deviations from slow roll.
We find that the condition is indeed satisfied even in situations consisting
of strong departures from slow roll, such as the punctuated inflationary
scenario.

\par

With the emergence of increasingly precise cosmological data, it has been
recognized that correlation functions beyond the power spectrum can act 
as powerful probes of the early universe.
However, as we had discussed in the introductory section, despite the 
relatively strong bounds that have been arrived at on the scalar non-Gaussianity 
parameter $\fnl$, there exist a wide range of models that remain consistent 
with the data.
The consistency relations can obviously play an important role to alleviate
the situation.
For instance, if the consistency relations can be observationally confirmed,
it can rule out many multi-field models of inflation and 
even, possibly, alternate scenarios such as the bouncing models (in this
context, see, for instance, Refs.~\cite{bm,ng-bm} and references therein).  
It seems worthwhile to closely investigate the conditions under which the 
consistency relation holds in, say, two-field models (see, for example, 
Ref.~\cite{cr-tfm}) and, in particular, examine in some detail the role the 
iso-curvature perturbations may play in this regard.
We are currently studying such issues. 


\section*{Acknowledgements}

The authors wish to thank J\'er\^ome Martin for collaborations and discussions 
as well as detailed comments on the manuscript. 
DKH wishes to acknowledge support from the Korea Ministry of Education, 
Science and Technology, Gyeongsangbuk-Do and Pohang City for Independent 
Junior Research Groups at the Asia Pacific Center for Theoretical Physics,
Pohang, Korea. 
The authors wish to acknowledge the use of Physics PC cluster at Pohang
University of Science and Technology, Pohang, Korea and the high performance 
computing facility at the Indian Institute of Technology Madras, Chennai, India. 



\begin{thebibliography}{99}
\bibitem{oa}
A.~A.~Starobinsky, JETP\ Lett.\ {\bf 30}, 682 (1979); V.~F.~Mukhanov and 
G.~V.~Chibisov, JETP\ Lett.\ {\bf 33}, 532 (1981); S.~W.~Hawking, Phys.\ 
Lett.\ B\ {\bf 115}, 295 (1982); A.~A.~Starobinsky, Phys.\ Lett.\ B\ 
{\bf 117}, 175 (1982); A.~Guth and S.-Y.~Pi, Phys.\ Rev.\ Lett.\ {\bf 49}, 
1110 (1982); A.~A.~Starobinsky, Astron.\ Lett.\ {\bf 9}, 302 (1983); 
V.~N.~Lukash, Sov.\ Phys.\ JETP {\bf 52}, 807 (1980); D.~H.~Lyth, Phys.\ Rev.\ 
D\ {\bf 31}, 1792 (1985).
\bibitem{texts}
E.~W.~Kolb and M.~S.~Turner, {\sl The Early Universe}\/ (Addison-Wesley, 
Redwood City, California, 1990); S.~Dodelson, {\sl Modern Cosmology}\/ 
(Academic Press, San Diego, U.S.A., 2003); V.~F.~Mukhanov, {\sl Physical 
Foundations of Cosmology}\/ (Cambridge University Press, Cambridge, 
England, 2005); S.~Weinberg, {\sl Cosmology}\/ (Oxford University Press, 
Oxford, England, 2008); R.~Durrer, {\sl The Cosmic Microwave Background}\/ 
(Cambridge University Press, Cambridge, England, 2008); D.~H.~Lyth and 
A.~R.~Liddle, {\sl The Primordial Density Perturbation}\/ (Cambridge 
University Press, Cambridge, England, 2009); P.~Peter and J-P.~Uzan, 
{\sl Primordial Cosmology}\/ (Oxford University Press, Oxford, England, 
2009); H.~Mo, F.~v.~d.~Bosch and S.~White, {\sl Galaxy Formation and 
Evolution}\/ (Cambridge University Press, Cambridge, England, 2010).
\bibitem{reviews} 
H.~Kodama and M.~Sasaki, Prog.\ Theor.\ Phys.\ Suppl.\ {\bf 78}, 1 (1984); 
V.~F.~Mukhanov, H.~A.~Feldman and R.~H.~Brandenberger, Phys.\ Rep.\ {\bf 215}, 
203 (1992); J.~E.~Lidsey, A.~Liddle, E.~W.~Kolb, E.~J.~Copeland, T.~Barreiro 
and M.~Abney, Rev.\ Mod.\ Phys.\ {\bf 69}, 373 (1997); 
A.~Riotto, arXiv:hep-ph/0210162; W.~H.~Kinney, astro-ph/0301448; J.~Martin, 
Lect. Notes Phys. {\bf 738}, 193 (2008); J.~Martin, Lect. Notes Phys. {\bf 669}, 
199 (2005); J.~Martin, Braz. J. Phys. {\bf 34}, 1307 (2004); B.~Bassett, 
S.~Tsujikawa and D.~Wands, Rev.\ Mod.\ Phys.\ {\bf 78}, 537 (2006); 
W.~H.~Kinney, arXiv:0902.1529 [astro-ph.CO]; L.~Sriramkumar, Curr.\ Sci.\ 
{\bf 97}, 868 (2009); D.~Baumann, arXiv:0907.5424v1 [hep-th].
\bibitem{cobe-1994}
C.~L.~Bennet {\it et al.},\/ Astrophys.\ J.\ Suppl.\ {\bf 436}, 423 (1994);
E.~L.~Wright {\it et al.},\/ Astrophys.\ J.\ Suppl.\ {\bf 436}, 443 (1994);
K.~M.~Gorski,\/ Astrophys.\ J.\ Suppl.\ {\bf 430}, L85 (1994);
K.~M.~Gorski {\it et al.},\/ Astrophys.\ J.\ Suppl.\ {\bf 430}, L89 (1994);
\bibitem{wmap-2009}
J.~Dunkley {\it et al.},\/ Astrophys.\ J.\ Suppl.\ {\bf 180}, 306 (2009);
E.~Komatsu {\it et al.},\/ Astrophys.\ J.\ Suppl.\ {\bf 180}, 330 (2009).
\bibitem{wmap-2011}
D.~Larson {\it et al.},\/ Astrophys.\ J.\ Suppl.\ {\bf 192}, 16 (2011);
E.~Komatsu {\it et al.},\/ Astrophys.\ J.\ Suppl.\ {\bf 192}, 18 (2011).
\bibitem{wmap-2013}
C.~L.~Bennett {\it et al.},\/ Astrophys.\ J.\ Suppl.\ {\bf 208}, 20 (2013); 
G.~Hinshaw {\it et al.},\/ Astrophys.\ J.\ Suppl.\ {\bf 208}, 19 (2013).
\bibitem{planck-2013-cmbps}
P.~A.~R.~Ade {\it et al.},\/ arXiv:1303.5075 [astro-ph.CO].
\bibitem{planck-2013-ccp}
P.~A.~R.~Ade {\it et al.},\/ arXiv:1303.5076 [astro-ph.CO].
\bibitem{planck-2013-ci}
P.~A.~R.~Ade {\it et al.},\/ arXiv:1303.5082 [astro-ph.CO].
\bibitem{bicep2-2014a}
P.~A.~R.~Ade {\it et al.},\/ arXiv:1403.4302 [astro-ph.CO].
\bibitem{bicep2-2014b}
P.~A.~R.~Ade {\it et al.},\/ Phys.\ Rev.\ Lett.\ {\bf 112}, 241101 (2014).
\bibitem{martin-2013}
J.~Martin, C.~Ringeval and V.~Vennin, arXiv: 1303.3787 [astro-ph.CO].
\bibitem{martin-2014a}
J.~Martin, C.~Ringeval, R.~Trotta and V.~Vennin, JCAP {\bf 1403}, 039 (2014); 
arXiv:1405.7272 [astro-ph.CO].
\bibitem{martin-2014b} 
J.~Martin, C.~Ringeval and V.~Vennin, arXiv:1407.4034 [astro-ph.CO].
\bibitem{fnl-komatsu}
E.~Komatsu and D.~N.~Spergel, Phys.\ Rev.\ D\ {\bf 63}, 063002 (2001).
\bibitem{planck-2013-cpng}
P.~A.~R.~Ade {\it et al.},\/ arXiv:1303.5084 [astro-ph.CO].
\bibitem{maldacena-2003}
J.~Maldacena, JHEP\ {\bf 0305}, 013 (2003).
\bibitem{creminelli-2004}
P.~Creminelli and M.~Zaldarriaga, JCAP {\bf 0410}, 006 (2004).
\bibitem{cr-rd}
C.~Cheung, A.~L.~Flitzpatrick, J.~Kaplan and L.~Senatore, JCAP {\bf 0802},
021 (2008); S.~Renaux-Petel, JCAP {\bf 1010}, 020 (2010); J.~Ganc and 
E.~Komatsu, JCAP {\bf 1012}, 009 (2010); P.~Creminelli, G.~D'Amico, 
M.~Musso and J.~Norena, JCAP {\bf 1111}, 038 (2011); D.~Chialva, JCAP 
{\bf 1210}, 037 (2012); K.~Schalm, G.~Shiu and T.~van der Aalst, JCAP 
{\bf 1303}, 005 (2013); E.~Pajer, F.~Schmidt and M.~Zaldarriaga, Phys.\
Rev.\ D\ {\bf 88}, 083502 (2013).
\bibitem{npfs}
L.~Senatore and M.~Zaldarriaga, JCAP {\bf 1208}, 001 (2012); P.~Creminelli, 
J.~Norena and M.~Simonovoc, JCAP {\bf 1207}, 052 (2012); P.~Creminelli, 
A.~Perko, L.~Senatore, M.~Simonovic and G.~Trevisan, JCAP {\bf 1311}, 015 (2013);
L.~Berezhiani and J.~Khoury, JCAP {\bf 1402}, 003 (2014); L.~Berezhiani, 
J.~Khoury and J.~Wang, arXiv:1401.7991 [hep-th]; H.~Collins, R.~Holman and 
T.~Vardanyan, arXiv:1405.0017 [hep-th].
\bibitem{tensor-bs}
J.~Maldacena and G.~L.~Pimentel, JHEP {\bf 1109}, 045 (2011); X.~Gao, 
T.~Kobayashi, M.~Yamaguchi and J.~Yokoyama, Phys.\ Rev.\ Lett.\ {\bf 107}, 
211301 (2011).
\bibitem{cc}
X.~Gao, T.~Kobayashi, M.~Shiraishi, M.~Yamaguchi, J.~Yokoyama and
S.~Yokoyama, arXiv:1207.0588 [astro-ph.CO].
\bibitem{cc2}
D.~Jeong and M.~Kamionkowski, Phys.\ Rev.\ Lett.\ {\bf 108}, 251301 (2012);
L.~Dai, D.~Jeong and M.~Kamionkowski, Phys.\ Rev.\ D\ {\bf 87}, 103006 (2013);
Phys.\ Rev.\ D\ {\bf 88}, 043507 (2013). 
\bibitem{sreenath-2013}
V.~Sreenath, R.~Tibrewala and L.~Sriramkumar, JCAP {\bf 1312}, 037 (2013).
\bibitem{kundu-2013}
S.~Kundu, arXiv:1311.1575 [astro-ph.CO].
\bibitem{sreenath-2014}
V.~Sreenath and L.~Sriramkumar, arXiv:1406.1609 [astro-ph.CO].
\bibitem{e-dfsr}
S.~M.~Leach and A.~R.~Liddle, Phys.\ Rev.\ D\ {\bf 63}, 043508 (2001);
S.~M.~Leach, M.~Sasaki, D.~Wands and A.~R.~Liddle, {\it ibid.} {\bf 64}, 
023512 (2001); R.~K.~Jain, P.~Chingangbam and L.~Sriramkumar, JCAP 
{\bf 0710}, 003 (2007).
\bibitem{cr-d-d-sr-ar}
R.~H.~Ribeiro, JCAP {\bf 1205}, 037 (2012); J.~Martin, H.~Motohashi and T.~Suyama, 
Phys.\ Rev.\ D\ {\bf 87}, 023514 (2013); M.~G.~Jackson and G.~Shiu, Phys.\ Rev.\ D\ 
{\bf 88}, 123511 (2013); R.~Flauger, D.~Green, R.~A.~Porto, JCAP {\bf 1308}, 032 
(2013); J.~Gong, K.~Schalm and G.~Shiu, Phys.\ Rev.\ D\ {\bf 89}, 063540 (2014).
\bibitem{cr-d-d-sr-nr}
P.~Adshead, W.~Hu, C.~Dvorkin and H.~V.~Peiris, Phys.\ Rev.\ D\ {\bf 84}, 
043519 (2011); A.~Achucarro, J-O.~Gong, G.~A.~Palma and S.~P.~Patil, Phys.\ 
Rev.\ D {\bf 87}, 121301 (2013).
\bibitem{dcr1}
M.~H.~Namjoo, H.~Firouzjahi and M.~Sasaki, Europhys.\ Lett.\ {\bf 101}, 39001 
(2013); X.~Chen, H.~Firouzjahi, M.~Namjoo and M.~Sasaki, Europhys.\ Lett.\ 
{\bf 102}, 59001 (2013).
\bibitem{dcr2}
J.~Ganc, Phys.\ Rev.\ D\ {\bf 84}, 063514 (2011); I.~Agullo and L.~Parker,
Phys.\ Rev.\ D\ {\bf 83}, 063526 (2011); Gen.\ Rel.\ Grav.\ {\bf 43}, 10 (2011).
\bibitem{bunch-1978}
T.~Bunch and P.~C.~W.~Davies, Proc.\ Roy.\ Soc.\ Lond.\ A\ {\bf 360}, 117 
(1978).
\bibitem{ng-ncsf}
D.~Seery and J.~E.~Lidsey, JCAP {\bf 0506}, 003 (2005); X.~Chen, Phys.\ 
Rev.\ D {\bf 72}, 123518 (2005); X.~Chen, M.-x.~Huang, S.~Kachru and 
G.~Shiu, JCAP {\bf 0701}, 002 (2007); D.~Langlois, S.~Renaux-Petel, 
D.~A.~Steer and T.~Tanaka, Phys.\ Rev.\ Lett.\ {\bf 101}, 061301 (2008); 
Phys.\ Rev.\ D\ {\bf 78}, 063523 (2008).
\bibitem{ng-reviews}
X.~Chen, Adv.\ Astron.\ {\bf 2010}, 638979 (2010); Y.~Wang, 
arXiv:1303.1523 [hep-th].
\bibitem{martin-2012a}
J.~Martin and L.~Sriramkumar, JCAP {\bf 1201}, 008 (2012).
\bibitem{hazra-2012}
D.~K.~Hazra, J.~Martin and L.~Sriramkumar, Phys.\ Rev.\ D\ {\bf 86}, 
063523 (2012). 
\bibitem{hazra-2013}
D.~K.~Hazra, L.~Sriramkumar and J.~Martin, JCAP {\bf 05}, 026 (2013).
\bibitem{p-law}
L.~F.~Abbott and M.~B.~Wise, Nucl.\ Phys.\ B\ {\bf 244}, 541 (1984); 
D.~H.~Lyth and E.~D.~Stewart, Phys.\ Lett.\ B\ {\bf 274}, 168 (1992); 
J.~Martin and D.~J.~Schwarz, Phys.\ Rev.\ D\ {\bf 57}, 3302 (1998); 
L.~Sriramkumar and T.~Padmanabhan, Phys. Rev. D 71, 103512 (2005).
\bibitem{gradshteyn-2007}
I.~S.~Gradshteyn and I.~M.~Ryzhik, {\sl Table of Integrals, Series and 
Products},\/ Seventh Edition (Acedemic Press, New York, 2007).
\bibitem{starobinsky-1992}
A.~A.~Starobinsky, Sov.\ Phys.\ JETP\ Lett.\ {\bf 55}, 489 (1992).
\bibitem{arroja-2011-2012}
F.~Arroja, A.~E.~Romano and M.~Sasaki, Phys.\ Rev.\ D\ {\bf 84}, 123503
(2011); F.~Arroja and M.~Sasaki, JCAP {\bf 1208}, 012 (2012).
\bibitem{martin-2014}
J.~Martin, L.~Sriramkumar and D.~K.~Hazra, arXiv:1404.6093 [astro-ph.CO]
\bibitem{gsr}
E.~D.~Stewart, Phys.\ Rev.\ D\ {\bf 65}, 103508 (2002); 
J.~Choe, J-O.~Gong and E.~D.~Stewart, JCAP {\bf 0407}, 012 (2004); 
J-O.~Gong, JCAP {\bf 0507}, 015 (2005).
\bibitem{pi}
R.~K.~Jain, P.~Chingangbam, J.-O.~Gong, L.~Sriramkumar and T.~Souradeep, 
JCAP {\bf 0901}, 009 (2009); R.~K.~Jain, P.~Chingangbam, L.~Sriramkumar 
and T.~Souradeep, Phys.\ Rev.\ D\ {\bf 82}, 023509 (2010). 
\bibitem{lp-ls}
L.~Lello, D.~Boyanovsky and R.~Holman, arXiv:1307.4066 [astro-ph.CO];
M.~Cicoli, S.~Downes and B.~Dutta, arXiv:1309.3412 [hep-th];
F.~G.~Pedro and A.~Westphal, arXiv:1309.3413 [hep-th].
\bibitem{l-22-40}
J.~A.~Adams, B.~Cresswell and R.~Easther, Phys.\ Rev.\ D\ {\bf 64}, 123514 
(2001); L.~Covi, J.~Hamann, A.~Melchiorri, A.~Slosar and I.~Sorbera, Phys.\ 
Rev.\ D\ {\bf 74}, 083509 (2006); J.~Hamann, L.~Covi, A.~Melchiorri and A.~Slosar, 
Phys.\ Rev.\ D\ {\bf 76}, 023503 (2007); M.~J.~Mortonson, M.~Joy, V.~Sahni and 
A.~A.~Starobinsky, Phys.\ Rev.\ D\ {\bf 77}, 023514 (2008); M.~Joy, A.~Shafieloo, 
V.~Sahni and A.~A.~Starobinsky, JCAP {\bf 0906}, 028 (2009); C.~Dvorkin, 
H.~V.~Peiris and W.~Hu, Phys.\ Rev.\ D\ {\bf 79}, 103519 (2009);
C.~Dvorkin and W.~Hu, Phys.\ Rev.\ D\ {\bf 81}, 023518 (2010);
W.~Hu, arXiv:1104.4500v1 [astro-ph.CO].
\bibitem{ci}
A.~Ashoorioon and A.~Krause,  arXiv:hep-th/0607001; A.~Ashoorioon, A.~Krause 
and K.~Turzynski, JCAP {\bf 0902}, 014 (2009).
\bibitem{hazra-2010}
D.~K.~Hazra, M.~Aich, R.~K.~Jain, L.~Sriramkumar and T.~Souradeep, JCAP 
{\bf 1010}, 008 (2010).
\bibitem{benetti-2011-13}
M.~Benetti, M.~Lattanzi, E.~Calabrese and A.~Melchiorri, Phys.\ Rev.\ D\ 
{\bf 84}, 063509 (2011); M.~Benetti, arXiv:1308.6406 [astro-ph.CO].
\bibitem{pso}
J.~Martin and C.~Ringeval, Phys.\ Rev.\ D\ {\bf 69}, 083515 (2004);
Phys.\ Rev.\ D\ {\bf 69}, 127303 (2004); JCAP {\bf 0501}, 007 (2005);
M.~Zarei, Phys.\ Rev.\ D\ {\bf 78}, 123502 (2008).
\bibitem{pahud-2009}
C.~Pahud, M.~Kamionkowski and A.~R.~Liddle, Phys.\ Rev.\ D\ {\bf 79}, 
083503 (2009).
\bibitem{flauger-2010}
R.~Flauger, L.~McAllister, E.~Pajer, A.~Westphal and G.~Xu, JCAP {\bf 1006}, 
009 (2010). 
\bibitem{aich-2013}
M.~Aich, D.~K.~Hazra, L.~Sriramkumar and T.~Souradeep, Phys.\ Rev. D\
{\bf 87}, 083526 (2013).
\bibitem{easther-2013}
H.~Peiris, R.~Easther and R.~Flauger, arXiv:1303.2616 [astro-ph.CO];
R.~Easther and R.~Flauger, arXiv:1308.3736 [astro-ph.CO].
\bibitem{meerburg-2013-14} 
P.~D.~Meerburg, D.~N.~Spergel and B.~D.~Wandelt, arXiv:1308.3704 [astro-ph.CO];
P.~D.~Meerburg and D.~N.~Spergel, arXiv:1308.3705 [astro-ph.CO]; P.~D.~Meerburg, 
arXiv:1406.3243 [astro-ph.CO].
\bibitem{be-fim}
J.~Martin, C.~Ringeval and R.~Trotta, Phys.\ Rev.\ D\ {\bf 83}, 063524
(2011); M.~J. Mortonson, H.~V.~Peiris and R.~Easther, Phys.\ Rev.\ D\ 
{\bf 83}, 043505 (2011); R.~Easther and H.~Peiris, Phys.\ Rev.\ D\ {\bf 85}, 
103533 (2012); J.~Norena, C.~Wagner, L.~Verde, H.~V.~Peiris and 
R.~Easther, Phys.\ Rev.\ D\ {\bf 86}, 023505 (2012).
\bibitem{rc}
D.~K.~Hazra, A.~Shafieloo and T.~Souradeep, JCAP {\bf 1307}, 031 (2013);
arXiv:1406.4827 [astro-ph.CO]; P.~Hunt and S.~Sarkar, arXiv:1308.2317 
[astro-ph.CO].
\bibitem{ne-ps}
D.~S.~Salopek, J.~R.~Bond and J.~M.~Bardeen, Phys.\ Rev.\ D\ {\bf 40}, 1753 
(1989); C.~Ringeval, Lect.\ Notes Phys.\ {\bf 738}, 243 (2008).
\bibitem{ng-ne} 
X.~Chen, R.~Easther and E.~A.~Lim, JCAP {\bf 0706}, 023 (2007); JCAP 
{\bf 0804}, 010 (2008).
\bibitem{ng-f} 
S.~Hotchkiss and S.~Sarkar, JCAP {\bf 1005}, 024 (2010); S.~Hannestad, 
T.~Haugbolle, P.~R.~Jarnhus and M.~S.~Sloth, JCAP {\bf 1006}, 001 (2010); 
R.~Flauger and E.~Pajer, JCAP {\bf 1101}, 017 (2011); P.~Adshead, W.~Hu, 
C.~Dvorkin and H.~V.~Peiris, Phys.\ Rev.\ D\ {\bf 84}, 043519 (2011); 
X.~Chen, JCAP {\bf 1201}, 038 (2012); P.~Adshead, W.~Hu and V.~Miranda, 
Phys.\ Rev.\ D\ {\bf 88}, 023507 (2013).
\bibitem{mayavi}
P.~Ramachandran and G.~Varoquaux, IEEE Computing in Science and Engineering 
{\bf 13}, 40 (2011)
\bibitem{bm}
D.~Battefeld and P.~Peter, arXiv:1406.2790 [astro-ph.CO].
\bibitem{ng-bm}
X.~Gao, M.~Lilley and P.~Peter, arXiv:1406.4119 [gr-qc].
\bibitem{cr-tfm}
V.~Assassi, D.~Baumann and D.~Green, JCAP {\bf 1211}, 047 (2012).
\end{thebibliography}
\end{document}